\documentclass[prd,onecolumn,nofootinbib,aps,tightenlines,preprintnumbers,12pt]{revtex4}

\usepackage{epsfig}
\usepackage{amsmath}
\usepackage{amsfonts}
\usepackage{dsfont}
\usepackage{fouridx}
\usepackage{hyperref}

  \newcount\hour \newcount\minute
  \hour=\time \divide \hour by 60
  \minute=\time
  \newcommand{\mydate}{\ \today \ - \number\hour :\ifnum \minute<10 0\fi
\number\minute}

\newcommand{\mpl}{M_{\rm pl}}

\def\tr{\mathop{\rm Tr}\nolimits}
\def\dim{\mathop{\rm dim}\nolimits}
\newcommand{\obs}{\widehat{\mathcal O}}
\newcommand{\Hilbert}{{\mathcal H}}

\newcommand{\order}[1]{\mathcal{O}(#1)}
\newcommand{\bra}[1]{\left\langle #1 \right|}
\newcommand{\ket}[1]{\left|#1\right\rangle}
\newcommand{\braket}[2]{\left\langle #1 \vphantom{#2} \right|
  \left. #2 \vphantom{#1} \right\rangle}
\newcommand{\bracket}[3]{\left\langle #1 \vphantom{#2#3} \right|
  #2 \left| #3 \vphantom{#1#2} \right\rangle}
\newcommand{\ketbra}[2]{\left|#1\left\rangle \right\langle #2 \right|}
\newcommand{\avg}[1]{\left\langle #1 \right\rangle}

\newcommand{\id}{{\mathds{1}}}
\newcommand{\scri}{\mathcal{I}}

\graphicspath{ {Figures/} }

\begin{document}

\preprint{\hbox{CALT-68-2878}  }

\title{De~Sitter Space Without Dynamical Quantum Fluctuations}

\author{Kimberly K. Boddy, Sean M. Carroll, and Jason Pollack}
\email{kboddy@theory.caltech.edu, seancarroll@gmail.com, jpollack@caltech.edu}
\affiliation{Physics Department, California Institute of Technology, Pasadena, CA 91125}

\begin{abstract}
We argue that, under certain plausible assumptions, de~Sitter space settles into a quiescent vacuum in which there are no dynamical quantum fluctuations.
Such fluctuations require either an evolving microstate, or time-dependent histories of out-of-equilibrium recording devices, which we argue are absent in stationary states.
For a massive scalar field in a fixed de~Sitter background, the cosmic no-hair theorem implies that the state of the patch approaches the vacuum, where there are no fluctuations.
We argue that an analogous conclusion holds whenever a patch of de~Sitter is embedded in a larger theory with an infinite-dimensional Hilbert space, including semiclassical quantum gravity with false vacua or complementarity in theories with at least one Minkowski vacuum.
This reasoning provides an escape from the Boltzmann brain problem in such theories.
It also implies that vacuum states do not uptunnel to higher-energy vacua and that perturbations do not decohere while slow-roll inflation occurs, suggesting that eternal inflation is much less common than often supposed.
On the other hand, if a de~Sitter patch is a closed system with a finite-dimensional Hilbert space, there will be Poincar\'e recurrences and dynamical Boltzmann fluctuations into lower-entropy states.
Our analysis does not alter the conventional understanding of the origin of density fluctuations from primordial inflation, since reheating naturally generates a high-entropy environment and leads to decoherence, nor does it affect the existence of non-dynamical vacuum fluctuations such as those that give rise to the Casimir effect.
\end{abstract}
\widetext
\maketitle
\vfill\eject

\tableofcontents
\vfill\eject

\baselineskip=14pt

\section{Introduction}

De~Sitter spacetime, and approximations to it, have come to play an important role in modern cosmology, both in inflation and in the likely future evolution of our universe.
The Hubble parameter in de~Sitter is constant and related to the cosmological constant by $H=\sqrt{\Lambda/3}$.
A stationary observer is surrounded by a cosmological horizon at a distance $R= H^{-1}$.
Quantum field theory (QFT) in curved spacetime describes a unique state that is both de~Sitter invariant and Hadamard (well-behaved at short distances), called the Euclidean (or Bunch-Davies~\cite{Bunch:1978yq,Bunch:1978yw}) vacuum for a free, massive scalar field or the Hartle-Hawking vacuum~\cite{Hartle:1976tp} for an interacting scalar field.
A particle detector sensitive to a field in the Hartle-Hawking vacuum will detect thermal Gibbons-Hawking radiation with a temperature $T=H/2\pi$~\cite{Gibbons:1977mu}.
Each horizon-sized patch (which we will henceforth simply call a ``patch'') of de~Sitter can be associated with an entropy equal to the area of the horizon in Planck units, $S = 3\pi/G\Lambda$.
In horizon complementarity, the quantum state of the bulk of each patch can be described by a density operator defined on a Hilbert space of dimension $\dim\Hilbert = e^S$~\cite{Banks:2000fe,Banks:2001yp}.

Conventional wisdom holds that the Hartle-Hawking vacuum experiences fluctuations, which may be thought of as either ``quantum'' or ``thermal,'' since a patch is a quantum system at a fixed temperature.
These fluctuations play several important roles in modern cosmological models.
During inflation, when the metric is approximately de~Sitter, fluctuations seed the density perturbations responsible for the cosmic microwave background (CMB) anisotropies and large-scale structure~\cite{Lyth:2009zz,Dodelson:2003ft,Baumann:2009ds}.
Eternal inflation (either stochastic~\cite{Vilenkin:1983xq,1987ZhETF..92.1137G,Goncharov:1987ir} or in a landscape of vacua~\cite{Weinberg:1987dv,Bousso:2000xa,Kachru:2003aw,Susskind:2003kw,Denef:2004ze}) makes use of fluctuations upward in energy density, often described as ``uptunneling''~\cite{Lee:1987qc,Aguirre:2011ac}.
Finally, the phenomena of Poincar\'e recurrences and fluctuations into Boltzmann brains can be problematic features of long-lived de~Sitter phases~\cite{Dyson:2002pf,Albrecht:2004ke,Bousso:2006xc}.

We will argue that some of this conventional wisdom is wrong.
Although a patch in the Hartle-Hawking vacuum is in a thermal state, we argue that it does not experience any kind of time-dependent fluctuations.
The density operator in the patch takes the form $\hat\rho \sim e^{-\beta\hat{H}}$, where $\beta = 1/T$ and $\hat H$ is the static Hamiltonian.
The state is stationary; there is no time dependence of any sort.
While it is true that an out-of-equilibrium particle detector inside the patch would detect thermal radiation, there are no such particle detectors floating around in the Hartle-Hawking vacuum.%
\footnote{See~\cite{Banks:2002wr} for a discussion on the difficulties of measurements in a finite-dimensional asymptotic de~Sitter space if a measuring device were indeed present.}
In fact, any particle detector placed in the vacuum would equilibrate, reaching a stationary state with thermal occupation numbers~\cite{Spradlin:2001pw}.

Fluctuations observed in a quantum system reflect the statistical nature of measurement outcomes.
Making a definite measurement requires an out-of-equilibrium, low-entropy detection apparatus that interacts with an environment to induce decoherence.
Quantum variables are not equivalent to classical stochastic variables.
They may behave similarly when measured repeatedly over time, in which case it is sensible to identify the nonzero variance of a quantum-mechanical observable with the physical fluctuations of a classical variable.
In a truly stationary state, however, there is no process of repeated measurements and hence no fluctuations that decohere.
We conclude that systems in such a state---including, in particular, the Hartle-Hawking vacuum---never fluctuate into lower-entropy states, including false vacua or configurations with Boltzmann brains.

Although our universe, today or during inflation, is of course not in the vacuum, the cosmic no-hair theorem~\cite{Wald:1983ky,Hollands:2010pr,Marolf:2010nz} implies that any patch in an expanding universe with a positive cosmological constant will asymptote to the vacuum.
Within QFT in curved spacetime, the Boltzmann brain problem is thus eliminated: a patch in eternal de~Sitter can form only a finite (and small) number of brains on its way to the vacuum.
At the same time, the standard story of inflationary perturbations remains intact: decoherence is accompanied by copious production of entropy during reheating.
Our analysis of fluctuations only calls into question the idea of dynamical transitions from stationary states to states with lower entropy.
We point out that the stochastic approximation in slow-roll eternal inflation~\cite{Vilenkin:1983xq,1987ZhETF..92.1137G,Goncharov:1987ir} makes use of such transitions to describe putative upward fluctuations of the inflation field.
Our picture rules out such fluctuations and may therefore change the conventional understanding of the conditions required for eternal inflation to occur.
In particular, eternal inflation is no longer an inevitable consequence of monomial inflation potentials like $V=m^2\varphi^2$.

The cosmic no-hair theorem is given in the context of QFT in curved spacetime.
Once quantum gravity is included, we need to be more careful.
If we accept the notion of horizon complementarity~\cite{Stephens:1993an,Susskind:1993if,Banks:2001yp,Parikh:2002py}, we should not use local QFT to simultaneously describe locations separated by a horizon.
Rather, we should treat each patch of eternal de~Sitter space, together with its horizon, as a closed, finite-dimensional quantum system.
The system is not stationary, so must undergo Poincar\'e recurrences as well as fluctuations, including into configurations we would describe as Boltzmann brains.
Alternatively, there might be a higher-entropy vacuum to which the system can decay, in which case the false de~Sitter vacuum patch can be thought of as an open subsystem embedded in a larger theory.
If the higher-entropy vacuum is de~Sitter, then the full system still has a finite-dimensional Hilbert space, subject to Poincar\'e recurrences and fluctuations.
If there is a Minkowski vacuum with potentially infinite entropy, the larger theory has an infinite-dimensional Hilbert space.
Here, we argue that the QFT analysis applies, and the patch rapidly approaches the vacuum and becomes quiescent, with only a finite number of fluctuations along the way.

This paper is organized as follows:
\begin{itemize}
\item In Section~\ref{fluctuations} we define what we mean by ``quantum fluctuations,'' distinguishing between three independent concepts: measurement-induced fluctuations, Boltzmann fluctuations, and vacuum fluctuations.
Measurement-induced fluctuations appear when an out-of-equilibrium measuring apparatus interacts with a quantum system, which results in time-dependent branching of the wave function.
In contrast, ``Boltzmann fluctuations'' are inherently dynamical statistical fluctuations, familiar from statistical mechanics.
``Vacuum fluctuations,'' which exist even in stationary states, represent differences between classical and quantum behavior, but do not correspond to dynamical (time-dependent) processes.
\item In Section~\ref{truevacuum} we examine eternal de~Sitter space in or near the unique Hartle-Hawking vacuum.
We first describe the system using QFT in a fixed background.
Because the Hartle-Hawking vacuum is stationary, we argue that there are no dynamical fluctuations, despite the fact that an out-of-equilibrium detector (of which there are none present) would measure a nonzero temperature.
The cosmic no-hair theorem ensures that all states evolve toward the vacuum, so the system must settle down to a state that is free of dynamical fluctuations.
In the context of horizon complementarity, however, each horizon volume can be treated as a system described by a finite-dimensional Hilbert space, and the cosmic no-hair theorem does not apply.
If de~Sitter space in horizon complementarity is eternal, there will be recurrences and Boltzmann fluctuations, and the conventional picture is recovered.
\item In Section~\ref{falsevacuum}, we turn to models that contain false de~Sitter vacua.
In semiclassical quantum gravity, or in complementarity in a landscape that includes a Minkowski vacuum, the dynamics occur in an infinite-dimensional Hilbert space.
The situation is then similar to QFT in global de~Sitter, where each patch can relax to a stationary quantum state, free of dynamical fluctuations.
In complementarity without a Minkowski vacuum, when all vacua are de~Sitter, there will still be Boltzmann fluctuations, since the total Hilbert space is finite-dimensional.
\item In Section~\ref{consequences}, we discuss the ramifications of this analysis.
First, the conventional Boltzmann brain problem is greatly ameliorated. Even with horizon complementarity, there are no fluctuations in the vacuum to lower-entropy states as long as the larger Hilbert space is infinite dimensional.
Similarly, we do not expect uptunneling to higher-energy vacua, which dramatically alters the picture of eternal inflation on a landscape.
The standard picture of density fluctuations from inflation remains unchanged, but the understanding of stochastic eternal inflation could be significantly different.
Finally, we note that these results depend crucially on one's preferred version of quantum mechanics.
\end{itemize}

\section{Fluctuations in Quantum Systems}
\label{fluctuations}

One way of thinking about the fluctuations of a quantum system is to consider an observable represented by a self-adjoint operator $\obs$.
If a state $\ket{\Psi}$ is not an eigenstate of $\obs$, then the variance
\begin{equation}
  (\Delta \obs)^2_\Psi = \langle\obs^2\rangle_\Psi - \langle\obs\rangle_\Psi^2
  \label{eq:QM-variance}
\end{equation}
will be strictly positive.
Hence, $\obs$ does not have a definite value.
However, a nonzero variance is not a statement about the \emph{dynamics} of the state, which may well be stationary; it is merely a statement about the distribution of measurement outcomes.
In quantum field theory, it is common to refer to radiative corrections from virtual particle pairs as ``quantum'', ``zero-point,'' or ``vacuum'' fluctuations, which give rise to phenomena such as the Lamb shift or Casimir effect;
they are not, however, ``fluctuations'' in the sense of a dynamical process that changes the state of the system.

In order to facilitate the investigation of the nature of fluctuations, we define and distinguish between the following types:
\begin{description}
\item[Vacuum fluctuations] are non-dynamical features of quantum states, which distinguish between classical and quantum behavior, and which ultimately arise as a consequence of the uncertainty principle.
  In quantum mechanics, vacuum fluctuations are described by \eqref{eq:QM-variance}; specifically, in quantum field theory, these fluctuations are understood as radiative corrections from virtual particle pairs.
\item[Measurement-induced fluctuations] are fluctuations whose dynamics are generated from a series of measurements of a quantum-mechanical system, resulting in decoherence and wave function branching.
\item[Boltzmann fluctuations] are dynamical fluctuations that arise when the microstate of a system is time-dependent, even though the coarse-grained macrostate may be stationary.
  They are associated, for example, with downward fluctuations in entropy of a thermal macrostate.
\end{description}
To study the Boltzmann brain problem and eternal inflation, we focus on the latter two types of fluctuations, in which time dependence plays a role.
For convenience we will use the term ``dynamical fluctuations'' to refer to either Boltzmann or measurement-induced types.
Vacuum fluctuations, of course, are a feature of all quantum systems, but our analysis will not be concerned with such non-dynamical features.

In the remainder of this section we clarify the meanings of ``measurement-induced fluctuations,'' along with the role of measuring devices, and ``Boltzmann fluctuations.''

\subsection{Decoherence and Everettian Worlds}
\label{decoherence}

Let us rehearse the standard understanding of quantum measurement and decoherence in the Everett formulation~\cite{Everett:1957hd,Schlosshauer:2003zy,wallace}.
There are two underlying postulates of the Everett formulation:
\begin{enumerate}
\item The world is represented by quantum states $\ket{\psi}$ that are elements of a Hilbert space $\Hilbert$.
\item The time evolution of states is generated by a self-adjoint Hamiltonian operator $\hat{H}$, according to the Schr\"odinger equation
  \begin{equation}
    \hat{H} \ket{\psi(t)} = i \partial_t \ket{\psi(t)} \ .
  \end{equation}
\end{enumerate}
In order to extract a description of a classical world from this formulation, we need to connect observables with measurement outcomes; this connection may be provided by the decoherence program.

As an illustration of decoherence, consider a Hilbert space that factors into an apparatus $A$ that may observe a system $S$:
\begin{equation}
  \Hilbert = \Hilbert_S \otimes \Hilbert_A \ .
\end{equation}
The Schmidt decomposition theorem allows us to write an arbitrary state as
\begin{equation}
  \ket{\Psi} = \sum_n c_n \ket{s_n} \ket{a_n} \ ,
  \label{eq:bipartite}
\end{equation}
where the $\ket{s_n}$ form an orthonormal basis for the system and $\ket{a_n}$ are orthogonal states of the apparatus.
We assume that $\dim\Hilbert_S < \dim\Hilbert_A$, and the sum over $n$ runs up to $\dim\Hilbert_S$.
The bipartite form of \eqref{eq:bipartite} is unique up to degeneracies in the coefficients $|c_n|$.
(For simplicity, we assume there are no degeneracies throughout the remainder of this paper.)
Although the Schmidt decomposition identifies a unique basis, there is no mechanism in place to ensure that system and apparatus states are ones that appropriately describe actual measurements.
Interactions between the system/apparatus and the environment are crucial for using decoherence to solve the measurement problem.

Incorporating the environment $E$, the Hilbert space is
\begin{equation}
  \Hilbert = \Hilbert_S \otimes \Hilbert_A \otimes \Hilbert_E \ .
  \label{hilbertsae}
\end{equation}
It may be possible to write a state in the full Hilbert space using a generalized Schmidt decomposition
\begin{equation}
  \ket{\Psi} = \sum_n c_n \ket{s_n} \ket{a_n} \ket{e_n} \ ,
  \label{tripartite}
\end{equation}
where $\ket{s_n}$ are system basis states; $\ket{a_n}$ are linearly independent, normalized apparatus states; and $\ket{e_n}$ are mutually noncollinear, normalized environment states.
The triorthogonal uniqueness theorem~\cite{Elby:1994} guarantees that the form of this tripartite decomposition, if it exists, is unique.
(Although this decomposition does not generically exist, it is a necessary feature of the standard decoherence program~\cite{Schlosshauer:2003zy}.)
Observations are restricted to the system and apparatus, so predictions of the outcomes of measurements are encoded in the reduced density matrix for the system and apparatus, found by tracing out the unobserved degrees of freedom of the environment from the full density matrix $\rho = \ketbra{\Psi}{\Psi}$:
\begin{align}
  \rho_{SA} &= \tr_E \ketbra{\Psi}{\Psi}\nonumber \\
  &= \sum_{m,n} c_m c_n^* \braket{e_n}{e_m}  \ket{s_m} \ket{a_m} \bra{s_n} \bra{a_n}\ .
  \label{reducedrho}
\end{align}

In order for this formalism to describe a quantum state that splits into independent Everettian branches or ``worlds,'' several requirements must be satisfied.
First, decoherence must occur---there must be no quantum interference between the different worlds, so observers on one branch evolve independently of the existence of other branches.
The absence of interference between states in $\Hilbert_S\otimes\Hilbert_A$ requires that the reduced density matrix~\eqref{reducedrho} be diagonal, {\it i.e.}, that the environment states associated with different branches be orthogonal.

Any density matrix is diagonal in some basis, but that basis might not be a physically viable one, nor one that is in the tripartite form of \eqref{tripartite}, where measurement outcomes are accurately reflected in the state of the apparatus.
The second requirement is, therefore, that there must exist a basis of apparatus ``pointer states'' in which decoherence naturally occurs through the dynamical diagonalization of $\rho_{SA}$ in this preferred basis~\cite{Zurek:1981xq,Zurek:1993ptp,Zurek:1998ji,Zurek:2003rmp,Schlosshauer:2003zy}.
A precise characterization of the pointer states is subtle and context-dependent but roughly corresponds to states of the apparatus that are macroscopically robust (stable).
Any interactions between the apparatus and environment should have a minimal effect on the system-apparatus correlations.
In principle, we can deduce the pointer states by writing the Hamiltonian as a sum of system/apparatus, environment, and interaction terms:
\begin{equation}
  \hat H = \hat H_{SA}\otimes \id_E + \id_{SA}\otimes \hat H_E + \hat H_I \ .
\end{equation}
The pointer states $\ket{a_n}$ are those whose projectors $\hat P_{n} = \ketbra{a_n}{a_n}$ commute with the interaction Hamiltonian,
\begin{equation}
  [\hat H_I,\hat P_{n}] = 0 \ .
\end{equation}
In practice, the fact that interactions are local in space implies that pointer states for macroscopic objects are those with definite spatial configurations.
For instance, if a large object (a billiard ball, a planet, a cat) is in a quantum superposition of two different position eigenstates, interactions with the environment (the air in a room, the cosmic background radiation) will rapidly cause those two possibilities to decohere, creating separate branches of the wave function.

The final feature that is important to describe branching is an arrow of time.
We conventionally imagine that worlds split via decoherence as time passes but almost never merge together, because we implicitly assume that the universe is very far from equilibrium and has evolved from a lower-entropy state in the past.
In the present context, ``low entropy'' means that subsystems begin in a particular state of little or no entanglement, as in \eqref{eq:unentangled}.
As we demonstrate in the next subsection, dynamical interactions between apparatus and environment naturally increase the amount of entanglement, leading to branching and generating entropy.%
\footnote{For the purposes of this paper, we are concerned with only the von~Neumann entropy from entanglements.
There is also the thermodynamic entropy associated with a mixed thermal density matrix, which sets an upper bound on the von~Neumann entropy.
As the quantum system thermalizes, the von~Neumann entropy approaches the thermodynamic entropy~\cite{Khlebnikov:2013cmsm}.}
The standard picture of decoherence and branching is specific to the far-from-equilibrium situation.
Near equilibrium, decoherence can arise through rare fluctuations, but is not tied to quantum measurements, as we discuss in Subsection~\ref{boltz-fluc}.

\subsection{Measurement-Induced Fluctuations}

We can use the decoherence program from the previous section to understand the nature of measurement-induced fluctuations.
For clarity in the following example, let us identify states in $S$, $A$, and $E$ explicitly with subscripts.
In the case of real-world quantum measurement, we posit that there is initially no entanglement between any of the factors:
\begin{equation}
  \ket{\Psi(t_0)} = \ket{\sigma_*}_S \ket{a_R}_A \ket{e_*}_E \ .
  \label{eq:unentangled}
\end{equation}
The initial state, denoted by an asterisk, of the system can be arbitrary; but the measuring apparatus must be in a specific ``ready'' state, denoted by the subscript $R$.
For definiteness, imagine that the system is a single qubit with basis states $\{\ket{+}_S, \ket{-}_S\}$.
The apparatus should begin in a ready state and record the results of repeated measurements of the system.
We take the apparatus state to be a tensor product of a number of registers (at least one for each measurement we want to perform), where each register is a qutrit with three basis states $\{\ket{+}_A, \ket{-}_A, \ket{0}_A\}$.
The ready state of the apparatus is $\ket{a_R}_A =\ket{000\cdots}_A$, and a measurement correlates one of the registers with the state of the system.
That is, under unitary evolution we record a measurement in the first register via
\begin{align}
  \ket{+}_S \ket{000\cdots}_A &\to \ket{+}_S \ket{+00\cdots}_A \ ,\\
  \ket{-}_S \ket{000\cdots}_A &\to \ket{-}_S \ket{-00\cdots}_A \ .
\end{align}
If the apparatus does not start in the ready state, we cannot be confident that it will end up correctly correlated with the state of the system.
Since unitary evolution must be reversible, there can be no valid evolution that takes $\ket{+}_S \ket{\psi}_A$ to $\ket{+}_S \ket{+}_A$ for every possible $\ket{\psi}_A$, for example.

Imagine that the system starts in a superposition, so the state takes the form
\begin{equation}
  \ket{\Psi(t_0)} = \left(\alpha \ket{+}_S + \beta \ket{-}_S \right)
  \ket{000\cdots}_A \ket{e_*}_E \ .
  \label{qf0}
\end{equation}
The first step in the evolution is premeasurement, which correlates the apparatus with the system:
\begin{equation}
  \ket{\Psi(t_1)} = \left(\alpha \ket{+}_S \ket{+00\cdots}_A
  + \beta \ket{-}_S \ket{-00\cdots}_A \right) \ket{e_*}_E \ .
  \label{qf1}
\end{equation}
The second step is decoherence, in which the apparatus becomes entangled with the environment:
\begin{equation}
  \ket{\Psi(t_2)} = \alpha \ket{+}_S \ket{+00\cdots}_A \ket{e_+}_E
  + \beta \ket{-}_S \ket{-00\cdots}_A \ket{e_-}_E \ .
  \label{qf2}
\end{equation}
Next, we reset in order to perform the measurement again, which means returning the system to its original state.
Generally, the environment states will also evolve during this operation.
We leave the apparatus unchanged in order to keep a record of the prior measurement outcomes:
\begin{equation}
  \ket{\Psi(t_3)} = \alpha\ket{\sigma_*}_S \ket{+00\cdots}_A \ket{\tilde e_+}_E
  + \beta \ket{\sigma_*}_S \ket{-00\cdots}_A \ket{\tilde e_-}_E \ .
  \label{qf3}
\end{equation}
Finally, we repeat the entire procedure, this time recording the measurement outcome in the second register of the apparatus.
After one more iteration of premeasurement and decoherence, we end up with
\begin{align}
  \ket{\Psi(t_4)} =& \
  \alpha^2 \ket{+}_S \ket{++0\cdots}_A \ket{e_{++}}_E\nonumber\\
  &+ \alpha\beta \ket{+}_S \ket{-+0\cdots}_A \ket{e_{-+}}_E\nonumber\\
  &+ \alpha\beta \ket{-}_S \ket{+-0\cdots}_A \ket{e_{+-}}_E\nonumber\\
  &+ \beta^2 \ket{-}_S \ket{--0\cdots}_A \ket{e_{--}}_E \ .
  \label{qf4}
\end{align}
At this point the wave function consists of four different decoherent branches, provided that all of the environment states are approximately orthogonal, $\braket{e_{\mu}}{e_\nu}_E \approx 0$.

In this context, the statement ``we observe quantum fluctuations'' is a statement about measurement-induced fluctuations: it is simply the observation that the history of each individual decoherent branch is one in which the state of the apparatus experiences a time series of observational outcomes, bouncing between $\ket{+}$ and $\ket{-}$.
On a randomly chosen branch, the history will exhibit fluctuations between the two outcomes, and all macroscopic objects are robust and physically well-defined (pointer states) by construction.
Schr\"odinger cat superpositions are not allowed, and different worlds or branches must evolve separately.

We see that obtaining the standard measurement outcomes requires both the apparatus to be initially in its ready state and the three Hilbert space factors (system/apparatus/environment) to be initially unentangled.
These conditions highlight the crucial role of entropy production in the branching of the wave function and thus in the existence of measurement-induced fluctuations.
The reduced density matrix $\rho_{SA}$ has a von~Neumann entropy
\begin{equation}
  S_{SA} = -\tr \rho_{SA} \log \rho_{SA} \ .
  \label{vnentropy}
\end{equation}
Since the state as a whole is pure in our example, all of the entropy comes from the entanglement between $SA$ and $E$.
In the initial state~\eqref{qf0}, there is no entanglement, and $S_{SA}=0$.
The entropy increases as the state evolves into two branches~\eqref{qf2} and again as it evolves into four branches~\eqref{qf4}.
Since the entropy of the pure state vanishes, the entropy of the environment equals that of the system/apparatus factor and increases as well.
Without entropy production, there are no measurement-induced fluctuations.

Now consider what happens if the entire wave function describing the system, apparatus, and environment ({\it i.e.}, the whole universe) begins in an energy eigenstate.
We assume there are interaction terms in the Hamiltonian that connect the different factors of the Hilbert space.
An energy eigenstate obeys
\begin{equation}
  \hat{H}\ket{\Psi_n} = E_n\ket{\Psi_n} \ ,
\end{equation}
where $\hat{H}$ is the full Hamiltonian.
Because the wave function is in an energy eigenstate, its time evolution just takes the form of multiplication by an overall time-dependent phase:
\begin{equation}
  \ket{\Psi_n(t)} = e^{-iE_n (t-t_0)}\ket{E_n} \ .
\end{equation}
The overall phase factor does not affect any of the observable properties of the state; therefore, it is sensible to refer to such a state as ``stationary,'' and its associated density operator
\begin{equation}
  \rho_\Psi = \ket{\Psi_n(t)}\bra{\Psi_n(t)} = \ketbra{E_n}{E_n}
\end{equation}
is manifestly time independent.
Another example of stationary density operator is that of a thermal state with temperature $\beta^{-1}$:
\begin{equation}
  \rho \sim \exp(-\beta \hat{H}) = \sum_n e^{-\beta E_n} \ketbra{E_n}{E_n} \ .
\end{equation}
Indeed, any density matrix diagonal in the energy eigenbasis will be stationary.

In a stationary state, none of the behavior we characterized as ``measurement-induced fluctuations''---branching of the wave function into a set of histories with stochastic measurement outcomes over time---is present.
In fact, there is no time dependence at all.%
\footnote{Even in stationary states, one can define an effective evolution with respect to correlations with a clock subsystem~\cite{Page:1983uc}.
The effective time parameter $\tau$ has nothing to do with the ordinary coordinate time $t$; all such time evolutions are present at every moment of (ordinary) time.
From this perspective, a large number of Boltzmann brains and similar fluctuations actually exist at every moment in an apparently stationary spacetime.
Such a conclusion would apply to Minkowski spacetime as well as to de~Sitter, in conflict with the conventional understanding that dynamical fluctuations in de~Sitter depend on the Gibbons-Hawking temperature (but see~\cite{Page:2005ur,Davenport:2010jy}).
This kind of effective evolution is fundamentally different from the ordinary evolution studied in this paper.}
Certainly, the variance of an observable $\hat{O}$ can be positive in a stationary state, but that variance only leads to dynamical fluctuations if the observable is actually measured.
Doing so requires an apparatus that is not itself stationary.
Indeed, the apparatus must start in a specific ready state, a condition that we may describe as low entropy.
If a quantum state describes the whole universe (as it does in cosmology), and this state is stationary, then it cannot undergo dynamical fluctuations, because nothing can actually change as time passes.
For a thermal state in particular, it will be the case that a particle detector beginning in its ready state would detect thermally fluctuating particles; but if all we have to use as a detector is a part of the stationary system itself, it will simply remain stationary, just as the rest of the quantum state does.

\subsection{Boltzmann Fluctuations}
\label{boltz-fluc}

There is an important difference between a quantum-mechanical thermal state and one in classical statistical mechanics.
Classically, a state in thermal equilibrium has a uniform temperature in space that is also constant in time.
However, this description is macroscopic and obtained by coarse graining.
Any realization of such a system with nonzero temperature has a microstate that is time-dependent.
For instance, the atoms and molecules in a box of gas are individually moving, even if the temperature and density are constant.
The system will, therefore, undergo rare fluctuations to nonequilibrium states.
The probability of observing a such a fluctuation to a state with entropy $\Delta S$ lower than equilibrium scales as $\sim e^{-\Delta S}$.
To avoid confusion we refer to such events, in which the evolution of the microstate causes a reduction in entropy, as ``Boltzmann fluctuations,'' to distinguish them from ``measurement-induced fluctuations'' where the wave function branches, which increase von~Neumann entropy.

In quantum mechanics, individual energy eigenstates are stationary, in contrast with classical states of nonzero energy.
Stationary quantum states will not experience Boltzmann fluctuations.
A statistical ensemble of stationary states will itself be stationary; we expect no Boltzmann fluctuations there as well. 
In particular, a closed system in a mixed thermal state, with a stationary density operator $\rho \sim e^{-\beta \hat{H}}$, should not have Boltzmann fluctuations when regarded as a statistical ensemble of energy eigenstates.
However, we most commonly encounter thermal density matrices after tracing over environmental degrees of freedom.
In that case the remaining system is not closed, and we need to be a bit more careful.

Consider a decomposition of a closed quantum system into a set of macroscopically observable system variables and an environment:
\begin{equation}
  \Hilbert = \Hilbert_S \otimes \Hilbert_E \ .
\end{equation}
(We have absorbed the apparatus that appears in \eqref{hilbertsae} into our definition of the macroscopic system.)
The environment includes local but microscopic variables (such as the positions and momenta of individual gas molecules, as opposed to macroscopic fluid variables such as temperature and pressure), as well as causally disconnected degrees of freedom (such as modes outside a cosmological horizon).
Expectation values of macroscopic observables in a pure state $\ket{\Psi} \in \Hilbert$ are encoded in the reduced density matrix $\rho_S = \tr_E|\Psi\rangle\langle\Psi|$, with entropy given by $S_S = -\tr \rho_S \log \rho_S$.
While the evolution of the pure state $\ket{\Psi}$ is unitary, that of $\rho_S$ is generally not.
It is described by a Lindblad equation~\cite{Lindblad:1975ef}, which allows for transfer of information between the macroscopic system and the environment:
\begin{equation}
  \dot\rho_S = i[\hat{H}_*,\rho_S] + \sum_n
  \left(\hat{L}_n \rho_S \hat{L}_n^\dagger
  - \frac{1}{2}\hat{L}_n^\dagger\hat{L}_n\rho_S
  - \frac{1}{2} \rho_S\hat{L}_n^\dagger\hat{L}_n\right) \ .
  \label{lindblad}
\end{equation}
The Lindblad operators $\hat{L}_n$ characterize the non-unitary part of the evolution of the system as induced by interactions with the environment, and will depend on the specific setup being studied.
The Hermitian operator $\hat{H}_*$ is not necessarily equal to the self-interaction Hamiltonian of the system alone; it captures the part of the entire Hamiltonian that induces unitary evolution on the reduced density matrix, including possible renormalization effects due to interaction with the environment.
A system far from equilibrium will generally exhibit dissipation and entropy increase (see \textit{e.g.}~\cite{2008arXiv0809.4403M}), and we may define a dissipation timescale on which the system will approach a stationary state.

\begin{figure}[t]
  \begin{center}
    \includegraphics[width=0.92\textwidth]{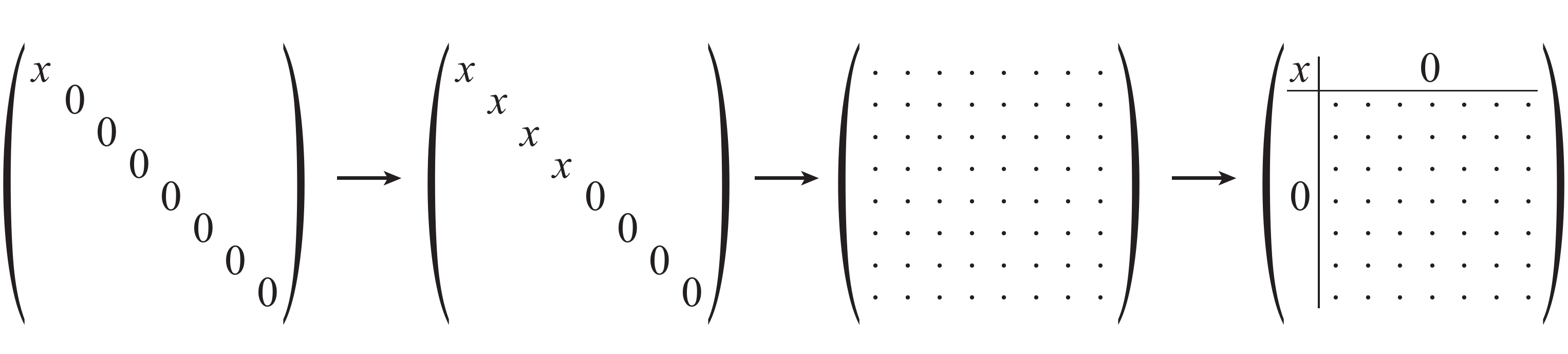}
  \end{center}
  \caption[Schematic evolution of a reduced density matrix in the pointer basis]{
    Schematic evolution of a reduced density matrix in the pointer basis.
    The density matrix on the left represents a low-entropy situation, where only a few states are represented in the wave function.
    There are no off-diagonal terms, since the pointer states rapidly decohere.
    The second matrix represents the situation after the wave function has branched a few times.
    In the third matrix, the system has reached equilibrium; the density matrix would be diagonal in an energy eigenbasis, but in the pointer basis, decoherence has disappeared and the off-diagonal terms are nonzero.
    The last matrix represents a Boltzmann fluctuation in which one pointer state has fluctuated into existence by decohering with respect to the other states.}
\label{equilibration-matrices}
\end{figure}

On much longer timescales, however, even systems with approximately-stationary reduced density matrices can experience decoherence and transitions to lower-entropy states, in precise analogy with Boltzmann fluctuations in classical statistical mechanics.
In Figure~\ref{equilibration-matrices} we provide a schematic representation of the evolution of the reduced density matrix, written in the pointer basis.
The first two entries show the splitting of branches of the wave function starting from a low-entropy configuration, as described for example by the transition from \eqref{qf0} to \eqref{qf2} and to \eqref{qf4}.
The state branches and decoheres, remaining diagonal in the pointer basis.
Eventually, it approaches equilibrium and becomes diagonal in the energy eigenbasis; by that point, the off-diagonal elements in the pointer basis are comparable to the diagonal ones, and the pointer states are no longer decoherent.
From equilibrium, there can be rare fluctuations (if the total Hilbert space is finite-dimensional) to lower-entropy configurations where one branch has once again decohered from the rest, as shown in the last entry.

Crucially, the existence of such fluctuations depends on the dimensionality $d_E$ of the Hilbert space $\Hilbert_E$ of the environment (assumed to be larger than the dimensionality of the system's Hilbert space $\Hilbert_S$).
For finite $d_E$, Hilbert space is bounded, and one can derive a quantum version of the Poincar\'e recurrence theorem~\cite{Dyson:2002pf}; for infinite $d_E$, the recurrence time goes to infinity, and excitations in the system can dissipate into the environment and never come back.
Zurek~\cite{Zurek:1982ii} has shown that, under reasonable assumptions concerning the initial wave function and the distribution of eigenvalues, the correlation amplitudes governing off-diagonal elements in the reduced density matrix will have an average of zero and experience fluctuations with a magnitude that scales as
\begin{equation}
  \Delta \sim d_E^{-1/2} \ .
  \label{eq:zurek_fluctuations}
\end{equation}
In a finite-dimensional Hilbert space, Boltzmann fluctuations are inevitable; however, in an infinite-dimensional space, the system can settle into equilibrium and stay there forever.
The reduced density matrix corresponding to the latter asymptotes to a stationary form, free of Boltzmann fluctuations.

This discussion presumes that the branching structure of the wave function can be discerned from the form of the reduced density matrix for the macroscopic variables $\Hilbert_S$.
In general, we cannot tell what states of a quantum system are actually realized on different branches simply by looking at its reduced density matrix.%
\footnote{We thank Alan Guth, Charles Bennett, and Jess Riedel for discussions on this point.}
For example, we might have a single qubit that takes on different states on three different branches of the wave function, specified by three mutually orthogonal environment states:
\begin{equation}
  \ket{\Psi} =
  \frac{1}{\sqrt{2}}\ket{+z}_S\ket{e_\uparrow}_E
  + \frac{1}{2}\ket{+x}_S\ket{e_\rightarrow}_E
  - \frac{1}{2}\ket{-x}_S\ket{e_\leftarrow}_E \ .
  \label{3ket1}
\end{equation}
The reduced density matrix for the qubit is
\begin{align}
  \rho_S &= \frac{1}{2}\ket{+z}_S\bra{+z}
  + \frac{1}{4}\ket{+x}_S\bra{+x}
  + \frac{1}{4}\ket{-x}_S\bra{-x} \\
 &=  \frac{3}{4}\ket{+z}_S\bra{+z}
  + \frac{1}{4}\ket{-z}_S\bra{-z} \ .
  \label{3rho2}
\end{align}
In the last line, the existence of three branches is completely obscured; the reduced density matrix does not reveal which states of the system exist as part of distinct worlds.

Thus, the reduced density matrix alone is not enough information to reveal what is truly happening inside a system.
Indeed, it is possible to construct a stationary reduced density matrix from an appropriate mixture of nonstationary states by tracing out the environment.
Therefore, the fact that a reduced density matrix is stationary does not suffice to conclude that there are no dynamical processes occurring on distinct branches within the system that it describes; for that, it is necessary to consider the full quantum state.
When we discuss the thermal nature of a patch of de~Sitter space in Section~\ref{eternaldS}, we have the benefit of knowing the full state of the de~Sitter vacuum, allowing us to circumvent this issue and draw conclusions about the (lack of) dynamics in a patch.

\section{Single de~Sitter Vacua}
\label{truevacuum}

We now apply these ideas to de~Sitter cosmology---specifically, to the case of a unique vacuum with $\Lambda > 0$.
In the Hartle-Hawking vacuum, the quantum state of any one causal patch is described by a thermal reduced density matrix.
As emphasized in Subsection~\ref{boltz-fluc} above, we cannot claim that the patch is stationary on the sole basis of its reduced density matrix; however, given that we know the full vacuum state, we argue that the patch is indeed stationary.
Were we to observe the patch, we would see fluctuations, but in the absence of an external observing device, nothing fluctuates.
In particular, there are no decohered branches of the wave function containing time-series records of fluctuating observables.
This picture does not apply if horizon complementarity is valid; in this case the entire Hilbert space is finite-dimensional, and unless it starts there, the state cannot asymptote to the vacuum as $t\rightarrow \infty$.
In complementarity, we expect Boltzmann fluctuations and Poincar\'e recurrences.

\subsection{Eternal de~Sitter}
\label{eternaldS}

Let us recall some basic properties of quantum fields in de~Sitter space~\cite{Birrell:1982ix,Spradlin:2001pw}.
De~Sitter space is the unique maximally symmetric spacetime with positive curvature.
In 4D, it has a scalar curvature $12 H^2$ and satisfies the Einstein equations with a cosmological constant $\Lambda=3H^2$, where $H^{-1}$ is the radius of de~Sitter space.
Consider a massive%
\footnote{We do not consider the massless case, since there is no (vacuum) state that is invariant under the full de~Sitter group~\cite{Allen:1985ux}, which is problematic for the cosmic no-hair theorem in Subsection~\ref{nohair}.
  However, if one assumes the shift invariance of the massless scalar field is just a global gauge transformation, then a fully de~Sitter invariant vacuum can in fact be defined~\cite{Page:2012fn}.},
noninteracting scalar field $\varphi$, which satisfies the Klein-Gordon equation
\begin{equation}
  (\Box - m^2) \varphi = 0
\end{equation}
in the de~Sitter metric.
In order to quantize fields in de~Sitter space, we must first choose a coordinate system.
There are numerous possibilities, but we narrow the scope to flat coordinates and static coordinates, as they are used most often in the literature.

In flat coordinates, the metric reads
\begin{equation}
  ds^2 = \frac{1}{H^2\tau^2} \left(-d\tau^2 + dx_i dx^i \right) \ ,
\end{equation}
which has the form of a flat, expanding Friedmann-Robertson-Walker metric with a constant Hubble parameter $H$ and conformal time $\tau$.
In these coordinates, there is no timelike Killing vector to provide a sensible prescription for defining modes of $\varphi$.
Since there is translational and rotational invariance among the spatial directions, we are still able to separate the mode solutions with wave number $\vec{k}$ as
\begin{equation}
  f(\tau) e^{i\vec{k}\cdot\vec{x}}
\end{equation}
for some function $f$.
Thus, we may attempt to define modes in the asymptotic regions of de~Sitter, $\scri^\pm$, by analogy with Minkowski space.
Because of this analogy, the vacuum defined by these modes will have the same symmetries as the free field Minkowski vacuum.
Unfortunately, the asymptotic regions are not static in an expanding universe, so we are left to define modes in the adiabatic approximation for a universe that has an infinitely slow expansion.
The Euclidean vacuum, formed from the adiabatic modes, is invariant under the de~Sitter group and, thus, does not change with time.
Although de~Sitter invariance alone does not define a unique state, the Euclidean vacuum is the unique de~Sitter-invariant Hadamard%
\footnote{Without the Hadamard condition~\cite{Candelas:1975du}, there are a continuum of de~Sitter-invariant states, known as the $\alpha$ vacua, which are related to one another via Bogoliubov transformations~\cite{Allen:1985ux}.}
state for a massive, noninteracting scalar field~\cite{Geheniau:1968bcs,Schomblond:1976xc,Chernikov:1968zm,Tagirov:1972vv,Mottola:1984ar,Allen:1985ux}.

In static coordinates the metric becomes
\begin{equation}
  ds^2 = -\left(1-H^2 r^2\right) dt^2 +
  \left(1-H^2 r^2\right)^{-1} dr^2 + r^2\,d\Omega^2 \ .
\end{equation}
These coordinates give a timelike Killing vector $-\partial_t$ that points toward the future (past) in the northern (southern) causal diamond, and we may use this Killing vector to define modes.
Following~\cite{Bousso:2001mw}, the mode expansions for the southern and northern diamonds of de~Sitter space are
\begin{align}
  \varphi^S &= \int_0^\infty d\omega\ \sum_{j=-\infty}^\infty
  \left[a_{\omega j}^S \varphi_{\omega j}^S +
  \left(a_{\omega j}^S\right)^\dagger
  \left(\varphi_{\omega j}^S\right)^* \right] \\
  \varphi^N &= \int_0^\infty d\omega\ \sum_{j=-\infty}^\infty
  \left[a_{\omega j}^N \varphi_{\omega j}^N +
  \left(a_{\omega j}^N\right)^\dagger
  \left(\varphi_{\omega j}^N\right)^* \right] \ ,
\end{align}
where $\omega$ is the mode frequency.
The operators $a_{\omega j}^N$ and $\left(a_{\omega j}^S \right)^\dagger$ are annihilation operators in the northern and southern diamonds.
The Euclidean vacuum is
\begin{equation}
  \ket{\Omega} = \prod_{\omega=0}^\infty \prod_{j=-\infty}^\infty
  \left(1-e^{-2\pi\omega}\right)^{1/2}
  \exp\left[e^{-\pi\omega}\left(a_{\omega j}^N \right)^\dagger a_{\omega j}^S \right]
  \ket{S} \otimes \ket{N} \ ,
  \label{omega}
\end{equation}
where $\ket{S}$ and $\ket{N}$ are the southern and northern no-particle vacua.
Ignoring gravitational back-reaction, the static Hamiltonian associated with the northern modes is
\begin{equation}
  \hat{H}_N = \int_0^\infty d\omega\ \sum_{j=-\infty}^\infty
  \left(a_{\omega j}^N\right)^\dagger a_{\omega j}^N\ \omega \ ,
\end{equation}
and the reduced density matrix in the northern diamond is
\begin{equation}
  \rho_N = \tr_S \ketbra{\Omega}{\Omega}
  = \left[\prod_\omega (1-e^{-2\pi\omega})\right] e^{-\beta \hat{H}_N} \ ,
 \label{rho_dS}
\end{equation}
which is a thermal density matrix with temperature $T=1/\beta$.

If the universe is in the Euclidean vacuum, the reduced density matrix describing the area inside a causal horizon is thermal.
In Subsection~\ref{boltz-fluc}, we argued that a subsystem with a thermal density matrix may still evolve into one with a Boltzmann fluctuation.
In the case of the Euclidean vacuum, however, we have both the reduced density matrix $\rho_N$ and the full quantum state $\ket{\Omega}$.
From an examination of \eqref{omega}, we see that the modes of a given frequency $\omega$ in the northern diamond are in a one-to-one correspondence with the modes in the southern diamond.
By tracing out the southern diamond to construct $\rho_N$, we know precisely which correlations we are discarding, mode by mode.
Furthermore, there is no interaction Hamiltonian between the northern and southern diamonds, since the diamonds are not in causal contact.
The entanglement structure is not disrupted by the separate evolution in each diamond, so dynamical processes, akin to the one shown in the last panel of Figure~\ref{equilibration-matrices}, are forbidden.
Then the reduced density matrix of each diamond is truly stationary, and no Boltzmann fluctuations are possible in either diamond.
(The spacetime geometry does not necessarily approach de~Sitter globally, but it asymptotes toward stationarity in each patch, which is all we really need.)

We have argued that there are no Boltzmann fluctuations in the de~Sitter vacuum.
It remains to determine whether the universe may actually be described by the de~Sitter vacuum.
Accordingly, the rest of our analysis consists of understanding the conditions under which the quantum state takes on this stationary vacuum form in different models.

\subsection{Cosmic No-Hair}
\label{nohair}

We turn now to situations, like that of our universe today, in which the universe is not in the vacuum but rather evolving in time.
We will see that, though there may be dynamical fluctuations initially if the state is very far from the vacuum, the state of a single patch will quickly approach the vacuum on time scales proportional to the inverse of the Hubble parameter, after which no fluctuations will arise.

We begin with the classical form of the cosmic no-hair theorem, which states that, given a positive vacuum energy density (\textit{i.e.}, a positive cosmological constant $\Lambda$), the metric evolves locally toward that of de~Sitter space~\cite{Wald:1983ky}.
Physically, excitations of de~Sitter (including matter and radiation fields with substantial energy densities) redshift away across the horizon, so every causal patch relaxes to the vacuum.

The physical intuition behind the cosmic no-hair theorem extends to quantum fields in curved spacetime.
For generic states, the expectation value of a massive scalar field $\varphi$ decays exponentially in time:
\begin{equation}
  \avg{\varphi(x)}_\psi = \order{e^{-M|\tau|}} \ ,
\end{equation}
for a decay constant $M>0$ and proper time $\tau$ between the point $x$ and some reference point at $\tau\to\infty$~\cite{Hollands:2010pr}.
Higher $n$-point correlation functions at large separations decay as well.
The vacuum is stable against perturbations and is an attractor state for local operators, whose expectation values in a generic state will approach the expectation values in the vacuum in the asymptotic region~\cite{Anderson:2000wx}.

A quantum-gravitational version of the no-hair theorem would presumably yield analogous results for the graviton field $h_{\mu\nu}$, but a scalar field can stand in as a proxy in order to make calculations manageable.
Although we have focused on a free scalar field theory to write an explicit form of the Euclidean vacuum and the reduced density matrix, the graviton has self interactions, so the analysis needs to be extended to an interacting scalar theory with a Hartle-Hawking vacuum.
For renormalizable interactions, the cosmic no-hair theorem still holds at an arbitrary number of loops, for arbitrary $n$-point functions, and for $D \geq 2$.
Furthermore, $M$ does not receive any radiative corrections.
The results of~\cite{Hollands:2010pr,Marolf:2010nz} show that the decay constant for massive%
\footnote{As previously mentioned, the massless case is problematic, since there is no de~Sitter-invariant vacuum in the noninteracting limit~\cite{Allen:1985ux}.
  With nonvanishing interactions, correlation functions of the field at large timelike separations grow no faster than a polynomial function of $H\tau$ at the perturbative level~\cite{Hollands:2011we}.
  There is, however, evidence at one and two loops that the 2-point correlation function decays as a polynomial of $H$~\cite{Hollands:2011we,Garbrecht:2011gu,Garbrecht:2013coa}.}
scalar fields is
\begin{equation}
  M=
  \begin{cases}
    \frac{3}{2}H
    & \textrm{for } m>\frac{3}{2}H \\
    \frac{3}{2}H - \sqrt{\frac{9}{4}H^2-m^2}
    & \textrm{for } 0<m\leq\frac{3}{2}H \ . \\
  \end{cases}
\end{equation}

If the universe is in an arbitrary state that is perturbed around the Hartle-Hawking vacuum, the state will approach the vacuum at large spacetime distances exponentially fast, with a decay constant $3H/2$ for large $m$.
Once the field correlations have sufficiently decayed, the arguments of Subsection~\ref{boltz-fluc} tell us that no dynamical fluctuations occur.

\subsection{Complementarity in Eternal de~Sitter}
\label{singledscomplementarity}

Horizon complementarity posits that the spacetime interpretation of a quantum state depends on the viewpoint of a specified observer~\cite{Stephens:1993an,Susskind:1993if,Banks:2001yp,Parikh:2002py}.
In particular, a description in terms of local quantum field theory will not extend smoothly beyond a horizon.
Applied to de~Sitter space, this philosophy implies that spacetime locality only applies within a cosmological horizon volume, and the corresponding quantum system has a finite-dimensional Hilbert space.
The Hilbert space of the patch can be decomposed as a product of bulk and boundary factors~\cite{Nomura:2011dt,Nomura:2011rb}:
\begin{equation}
  \Hilbert = \Hilbert_{\textrm{bulk}}\otimes \Hilbert_{\textrm{boundary}} \ .
  \label{hilbertcomp}
\end{equation}
(We ignore a possible factor corresponding to singular spacetime geometries, which will not be important for our analysis.)

From the Bekenstein-Hawking relation~\cite{Bekenstein:1973ur,Hawking:1971tu}, the entropy associated with the patch is one quarter of the area of the horizon: $S_{\rm dS} = \mathcal{A}/4$.
This entropy is related to the density matrix $\rho\sim e^{-\beta \hat{H}}$ for the patch via $S_{\rm dS} = -\tr_{\textrm{boundary}} \rho\ln\rho$, so the patch is thermal even if the system as a whole is in a pure quantum state.
The energy spectrum is discrete, with only a finite number of eigenvalues with energies less than any given cutoff value~\cite{Goheer:2002vf}.

If we interpret the entropy as being the logarithm of the number of quantum states, the horizon patch is analogous to a closed thermal system at a temperature $T$~\cite{Banks:2000fe,Banks:2001yp}.
Although the relationship $\dim\Hilbert = e^S$ holds only at infinite temperature, there are compelling reasons (\textit{e.g.}, from black holes) to think that the static Hamiltonian is bounded from above~\cite{Banks:2005bm}.%
\footnote{For subtleties involving the use of the static Hamiltonian in quantum gravity, see~\cite{Giddings:2007nu}.}
In our discussion of complementarity, we assume that this bound exists and that the dimension of the Hilbert space
\begin{equation}
  \dim\Hilbert = e^{2S_{\rm dS}} = \exp(6\pi\Lambda^{-1})
\end{equation}
is finite.
(The factor of $2$ comes from the fact that the bulk and boundary components have equal dimensionality.)

\begin{figure}[t]
  \begin{center}
    \includegraphics[width=0.42\textwidth]{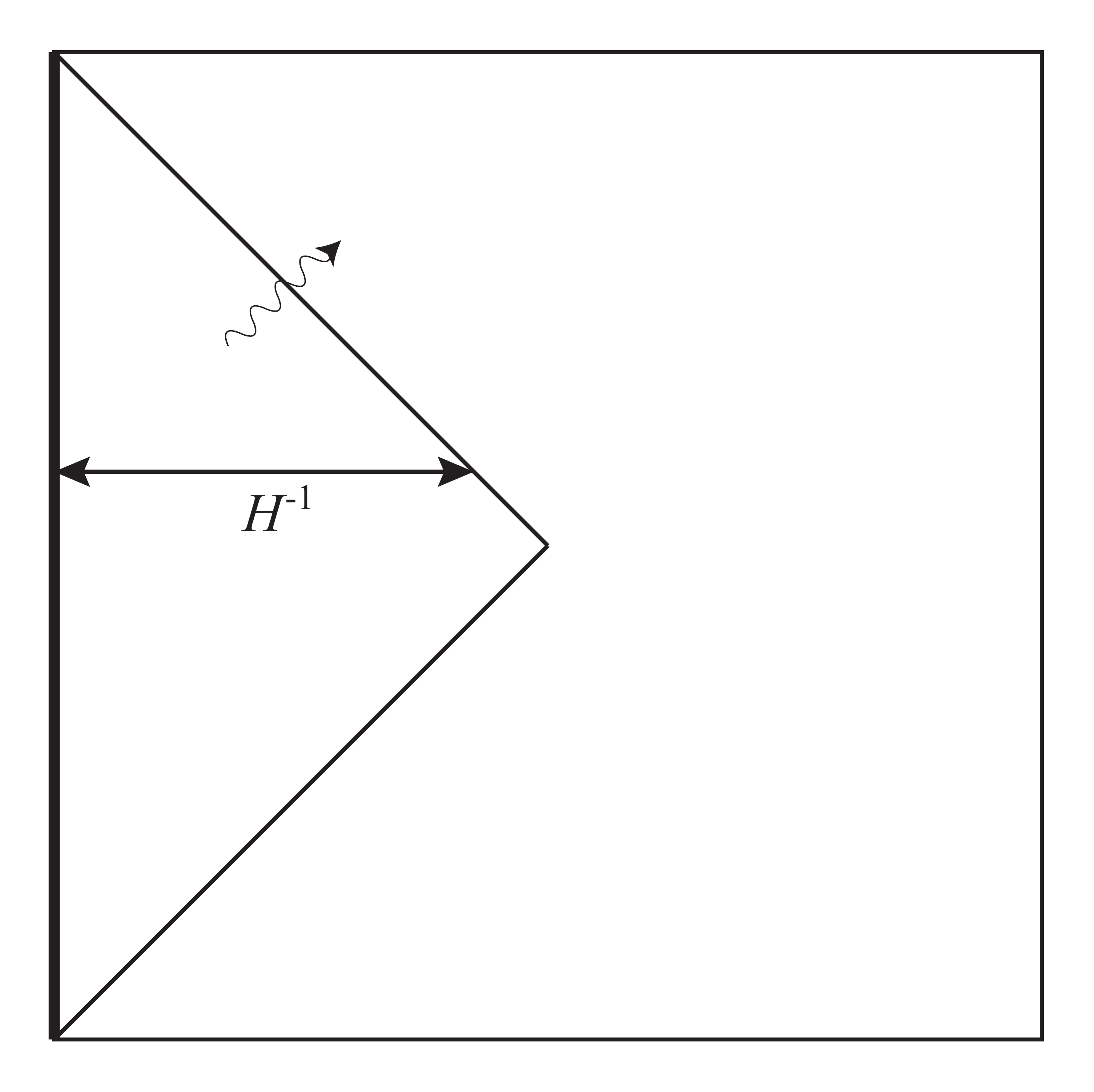}
    \includegraphics[width=0.42\textwidth]{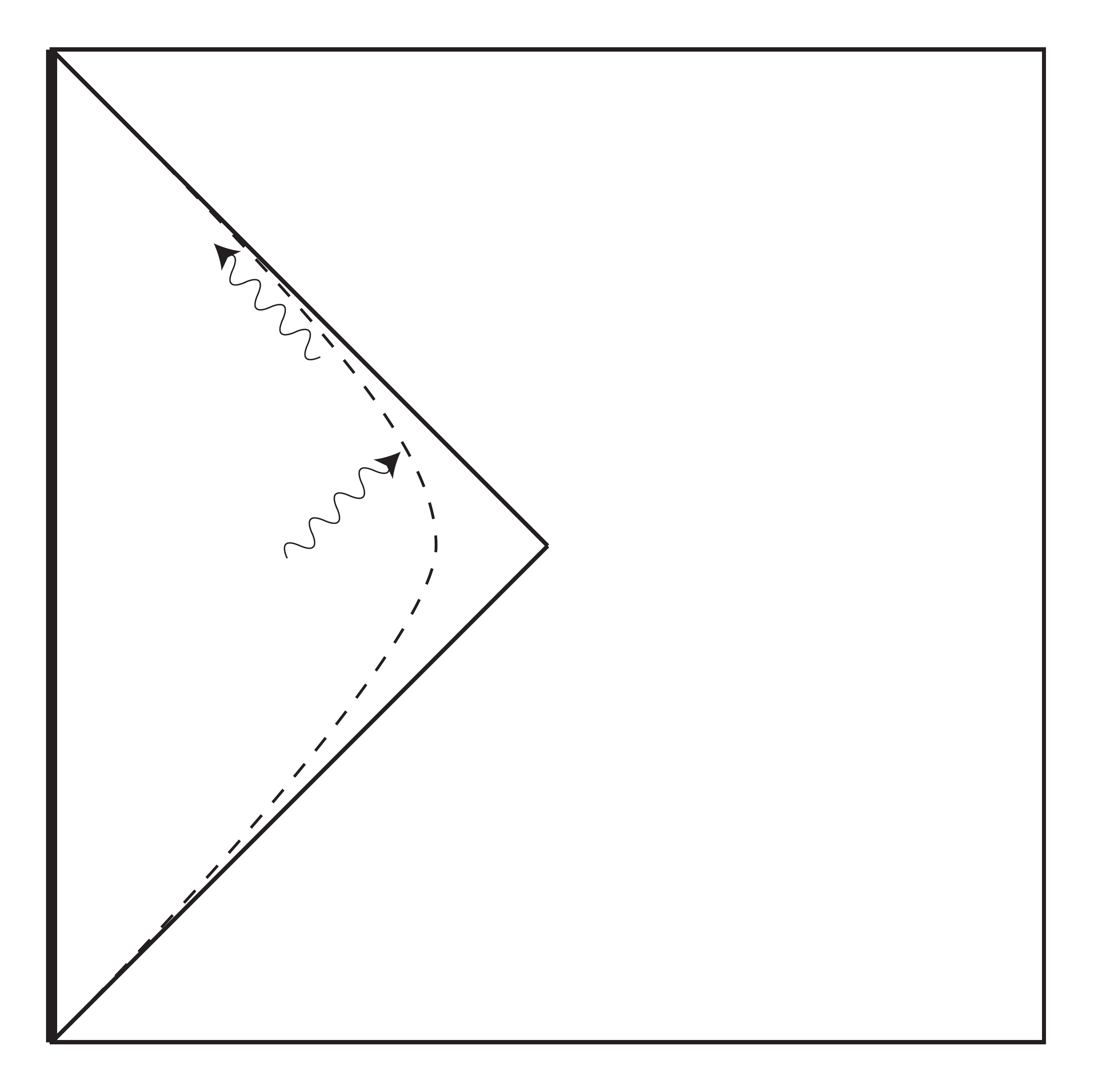}
  \end{center}
  \caption[Conformal diagrams for de~Sitter space]{
    Conformal diagrams for de~Sitter space in the global (QFT) picture [left] and with horizon complementarity [right].
    We consider an observer at the north pole, represented by the line on the left boundary and their causal diamond (solid triangle).
    The wavy line represents excitations of the vacuum approaching the horizon.
    In QFT in curved spacetime, the excitation exits and the state inside the diamond approaches the Hartle-Hawking vacuum, in accordance with the cosmic no-hair theorems.
    In contrast, horizon complementarity implies that excitations are effectively absorbed at the stretched horizon (dashed curve just inside the true horizon) and eventually return to the bulk.}
\label{conformal-fig1}
\end{figure}

The complementarity picture of eternal de~Sitter with a unique vacuum state is, therefore, very different from the situation of QFT in a de~Sitter background discussed in Subsection~\ref{eternaldS}.
In the latter, the ability of excitations to leave the horizon and never return depended crucially on the fact that Hilbert space was infinite-dimensional.
In complementarity, eternal de~Sitter space is a truly closed finite-dimensional system, subject to Poincar\'e recurrences~\cite{Dyson:2002pf}.
Of course, there is a true vacuum state, the lowest-energy eigenstate, that is strictly stationary, but a generic state is nonstationary.
We may think of excitations as being absorbed by a stretched horizon with a finite area and eventually being emitted back into the bulk, as shown in Figure~\ref{conformal-fig1}.
Boltzmann fluctuations into lower-entropy states (described in Subsection~\ref{boltz-fluc}) are allowed, in agreement with the conventional picture of a thermal de~Sitter patch.
As we argue below, this story changes in important ways in theories with more than one metastable vacuum.

\section{Multiple Vacua}
\label{falsevacuum}

In this section we consider theories with more than one metastable potential minimum, at least one of which has $\Lambda >0$, as portrayed schematically in Figure~\ref{vacua-fig}.
We consider the existence of dynamical fluctuations in both the lowest-energy ``true'' vacuum and in any higher-energy false vacua.
For convenience, we limit our attention to vacua with non-negative energy, $\Lambda \geq 0$.
Transitions from vacua with $\Lambda \geq 0$ to those with $\Lambda < 0$ generally result in singular crunches; evolution might continue via quantum-gravity effects, but we will not address that possibility here.

\subsection{Semiclassical Quantum Gravity}
\label{semiclassical}

\begin{figure}[t]
  \begin{center}
     \includegraphics[width=0.7\textwidth]{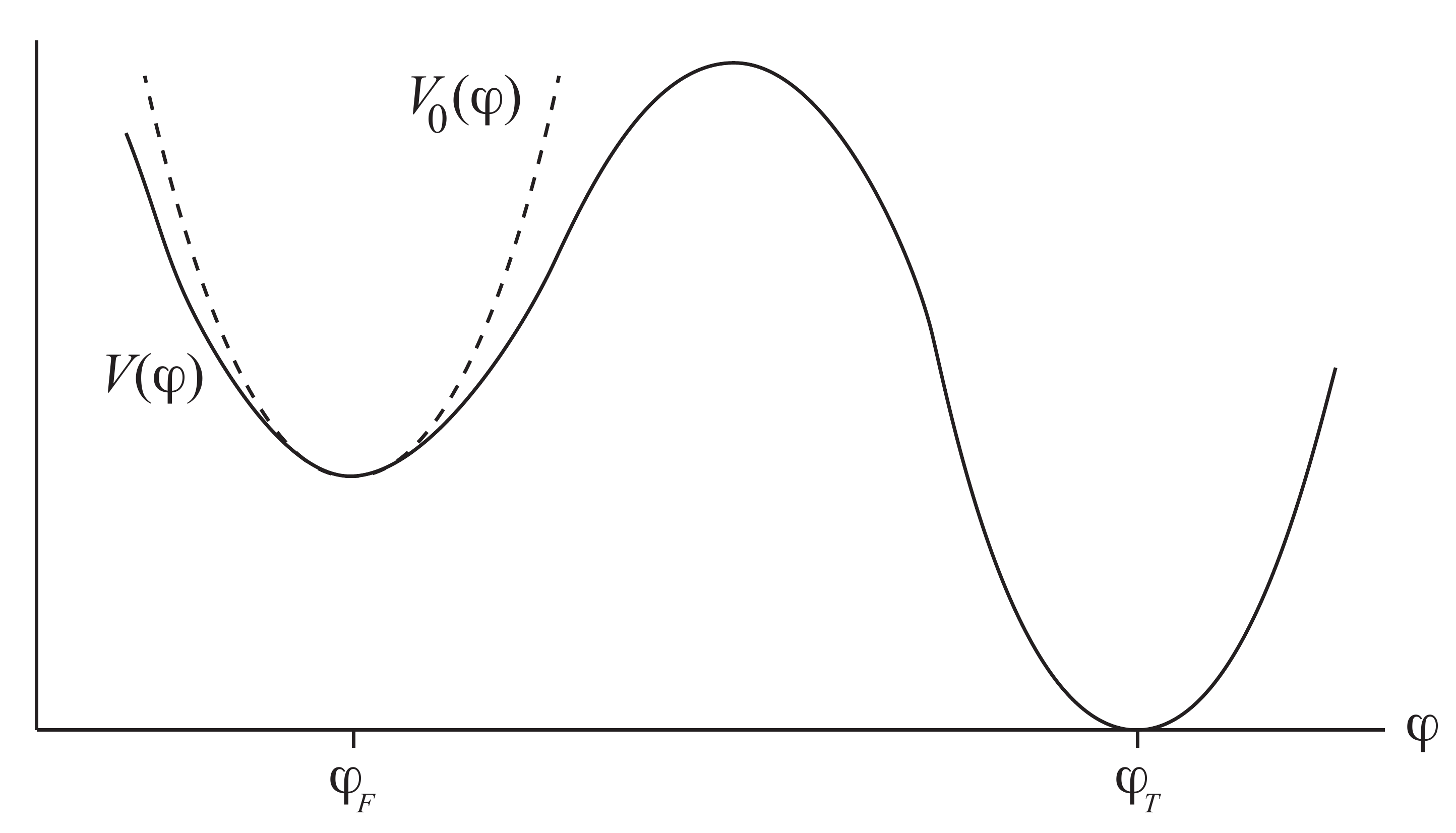}
  \end{center}
  \caption[Scalar field potential with multiple local minima]{
    A scalar field potential with multiple local minima.
    The global minimum corresponds to the true-vacuum value $\varphi_T$ (which may have $\Lambda =0$ or $\Lambda >0$), and for simplicity we have portrayed a single false-vacuum value $\varphi_F$.
    The dashed line represents the perturbative Hamiltonian for the false vacuum, in which the potential is given by a local approximation to the true potential in the vicinity of $\varphi_F$.}
\label{vacua-fig}
\end{figure}

We first consider semiclassical quantum gravity, by which we mean QFT coupled to a classical (but dynamical) spacetime background.
Coleman studied false vacua in this context and calculated the rate at which a higher-energy vacuum would decay to a lower-energy state via bubble nucleation~\cite{Coleman:1977py,Coleman:1980aw}.
It is useful to consider an analogous problem in one-dimensional quantum mechanics, in which a single particle moves in a potential $V(x)$, with a global (true) minimum at $x_T$ and a local (false) minimum at $x_F$.
Then, one can calculate the transition amplitude using the path integral defined with respect to Euclidean time $T$:
\begin{equation}
  \bracket{x_T}{e^{-H T}}{x_F} = N \int [dx] e^{-S_E[x(T)]} \ ,
  \label{vacuum-transition}
\end{equation}
where $H$ is the Hamiltonian and $S_E$ is the Euclidean action, while the states $\ket{x_T}$ and $\ket{x_F}$ are position eigenstates.
This quantity can be calculated using instanton methods and represents the amplitude for finding the particle at position $x_T$, given that it started at position $x_F$---something that might be of relevance to an observer measuring the position of the particle.
An analogous field theory calculation can be used to calculate the rate of from one field configuration $\ket{\varphi_1(x)}$ to another $\ket{\varphi_2(x)}$, including the tunneling rate from one vacuum to another, as shown in Figure~\ref{vacua-fig}.

Our interest, however, is not in what an out-of-equilibrium observer with a field-value detection device would measure, but in how quantum states evolve in isolated patches of de~Sitter space.
Eigenstates of the field operator $\hat\varphi(x)$ are not energy eigenstates; therefore, we need to be careful when we use terms such as ``false vacuum'' and ``true vacuum'' to refer to quantum states rather than field values.
For some purposes it is useful to study eigenstates of a perturbative Hamiltonian constructed by approximating the potential in the vicinity of one local minimum, as shown for $\varphi_F$ in Figure~\ref{vacua-fig}.
In that case the results from Section~\ref{eternaldS}, where we studied QFT in a fixed de~Sitter background, are relevant.

Consider first the true vacuum quantum state $\ket{0}$ of the full theory.
A generic homogeneous field value $\varphi_*$ will have some nonzero overlap with this state, $\braket{\varphi_*}{0} \neq 0$, but the field will be mostly localized near the global minimum value $\varphi_T$.
While it is difficult to rigorously prove a version of the cosmic no-hair theorem for this interacting theory, we intuitively expect the physics in this case to mirror that of QFT with a unique de~Sitter vacuum.
Namely, excitations above the lowest-energy state will dissipate outside the horizon, and each local patch will approach the vacuum state $\ket{0}$.
This state is stationary, and we expect no measurement-induced or Boltzmann fluctuations.
Since we are dealing with QFT, the Hilbert space is infinite-dimensional, and there are no recurrences.

We also do not expect uptunneling to a higher-energy vacuum from the true vacuum state for the same reason (energy eigenstates are stationary and do not fluctuate).
This assertion might seem to be in tension with the existence of instantons that contribute a nonzero amplitude to processes analogous to \eqref{vacuum-transition}, but such a counterargument confuses field values with quantum states.
Although there are instanton solutions, their role is to shift the value of the vacuum energy in the true vacuum from what one would compute in a local approximation to the effective potential near $\varphi_T$.
As noted above, the nonzero overlap between two perturbative vacua can be interpreted as a transition rate between them.
But we are interested in the states of definite semiclassical geometry, which should correspond to vacua of the full potential, where instanton corrections have already been taken into account.

The situation is analogous to that of the QCD vacuum, where instantons connecting vacua of different winding numbers provide a shift in energy that depends on the value of $\theta_{\rm QCD}$.
The QCD vacuum is a single, static state which incorporates the instanton corrections, not constantly occurring dynamical transitions between states of definite winding number, just like a harmonic oscillator in an energy eigenstate is static rather than undergoing constant fluctuations.
Even though we can write the QCD vacuum in a basis of states of different winding number, or an energy eigenstate of the harmonic oscillator in a basis of position eigenstates, the lesson of Section \ref{fluctuations} is that such descriptions have no physical reality.
Instantons are important for calculating energy eigenvalues, but once the quantum system is in a stationary state such as the vacuum $\ket{0}$, they do not describe true dynamical transitions.
The local perturbative vacuum will be unstable to uptunneling via instantons, but that's not the true nonperturbative vacuum into which the system settles.

Next we turn to false vacua.
A semiclassical state with $\avg{\varphi} = \varphi_F$ is not strictly a vacuum state, or indeed any form of energy eigenstate, as it will decay via tunneling.
We may nevertheless consider the energy eigenstates of the perturbative Hamiltonian, obtained by locally approximating the potential in the vicinity of $\varphi_F$, as shown in Figure~\ref{vacua-fig}.
These are not energy eigenstates of the full Hamiltonian, but their dynamics are well-described by a combination of processes near the false-vacuum value plus decays via bubble nucleation.
We may think of the ``false de~Sitter vacuum state'' as the Hartle-Hawking vacuum state of this perturbative Hamiltonian.
Once again, we expect excitations to rapidly dissipate by leaving the horizon, resulting in a state that does not exhibit thermal fluctuations.
We refer to such states as ``quiescent'' (reserving the term ``stationary'' for true energy eigenstates).

We are left with two kinds of possible non-perturbative processes to consider: downtunneling to lower-energy vacua and uptunneling to higher-energy vacua.
First, we examine downtunneling.
In the conventional picture of false-vacuum decay, a small bubble of true vacuum nucleates and grows at nearly the speed of light.
This picture is clearly a semiclassical description of a single branch of the wave function, rather than a full treatment of the quantum state.
We can decompose the Hilbert space into the product of the state space of a smooth background field $\varphi_\lambda(x)$ and small-scale fluctuations:
\begin{equation}
  \Hilbert = \Hilbert_{\varphi_\lambda}\otimes\Hilbert_{\delta\varphi} \ .
\end{equation}
Here, $\lambda$ is a length scale used to smooth the field.
The factor $\Hilbert_{\varphi_\lambda}$ includes configurations with bubbles of different sizes and locations, as well as completely homogeneous configurations.
When a bubble nucleates, some of the energy density that was in the potential for $\varphi$ gets converted into fluctuation modes, resulting in the production of entropy.
Therefore, a reduced density matrix for the background field obtained by tracing over $\Hilbert_{\delta\varphi}$ will exhibit decoherence, as the fluctuations produced by bubbles in different locations will generically be orthogonal to each other.
In that sense, the semiclassical configurations described by bubble nucleation correspond to truly distinct branches of the wave function.
With that single caveat, we agree with the standard picture of downtunneling to lower-energy vacua.

\begin{figure}[t]
  \begin{center}
    \includegraphics[width=0.32\textwidth]{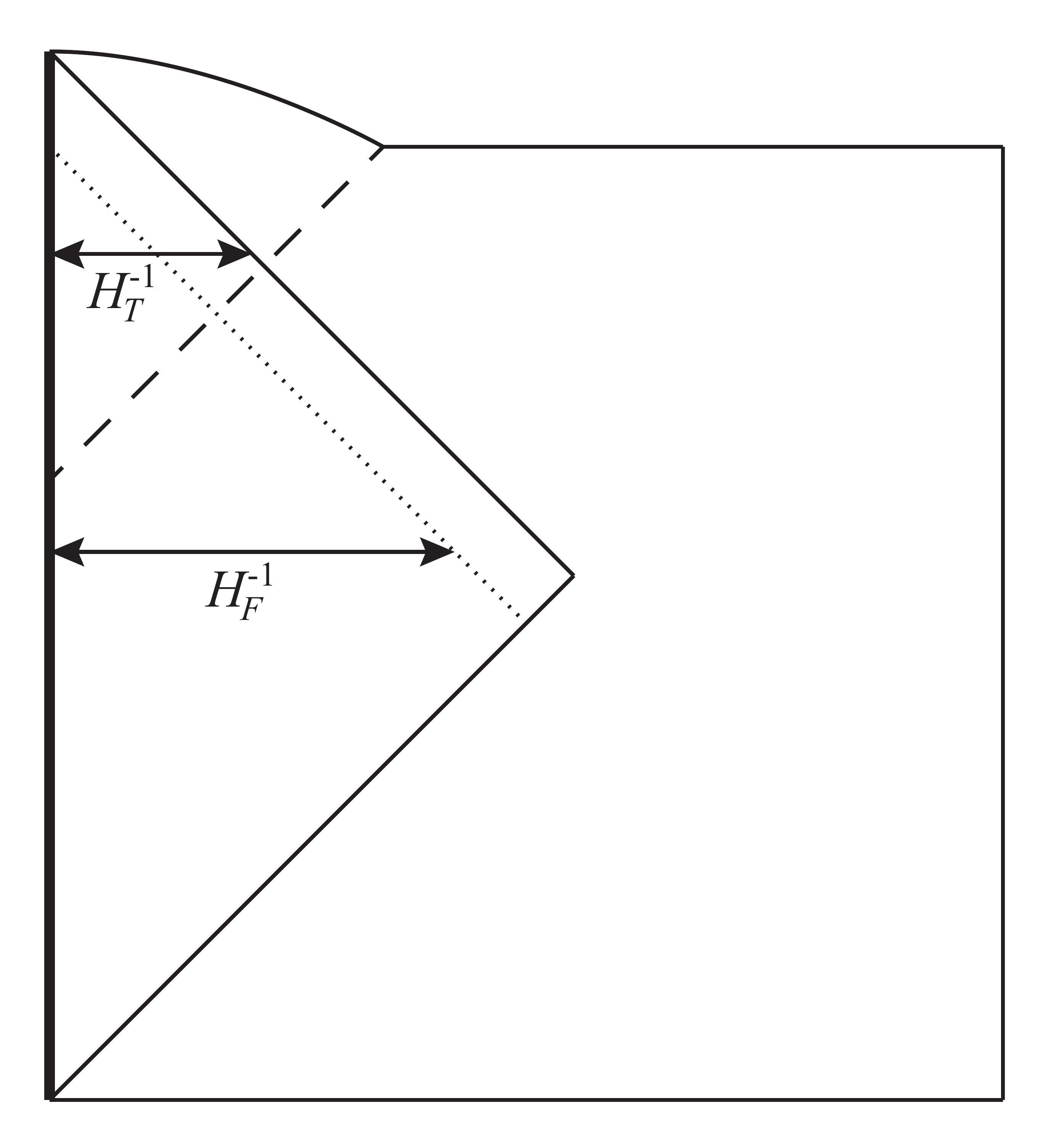}
    \includegraphics[width=0.32\textwidth]{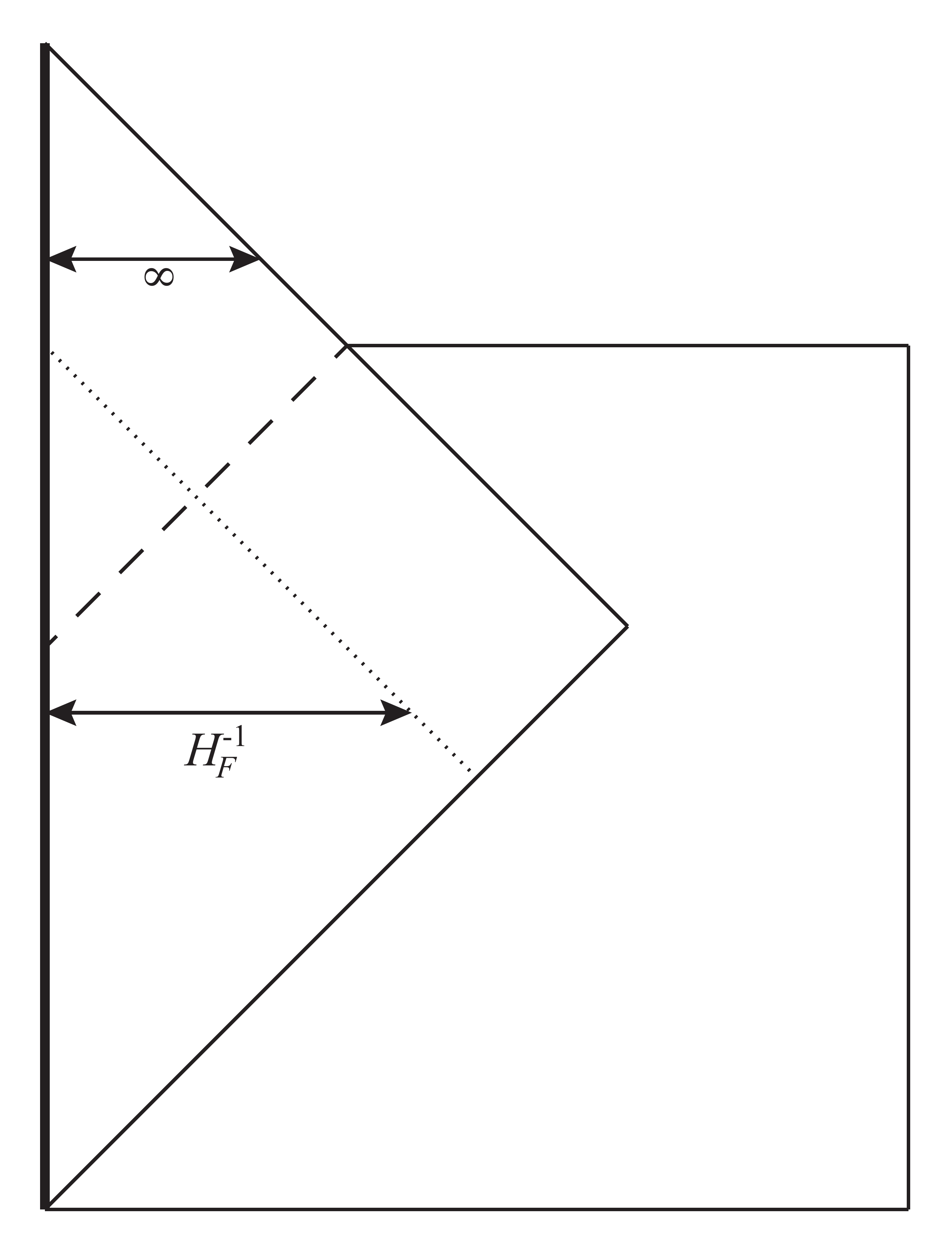}
    \includegraphics[width=0.32\textwidth]{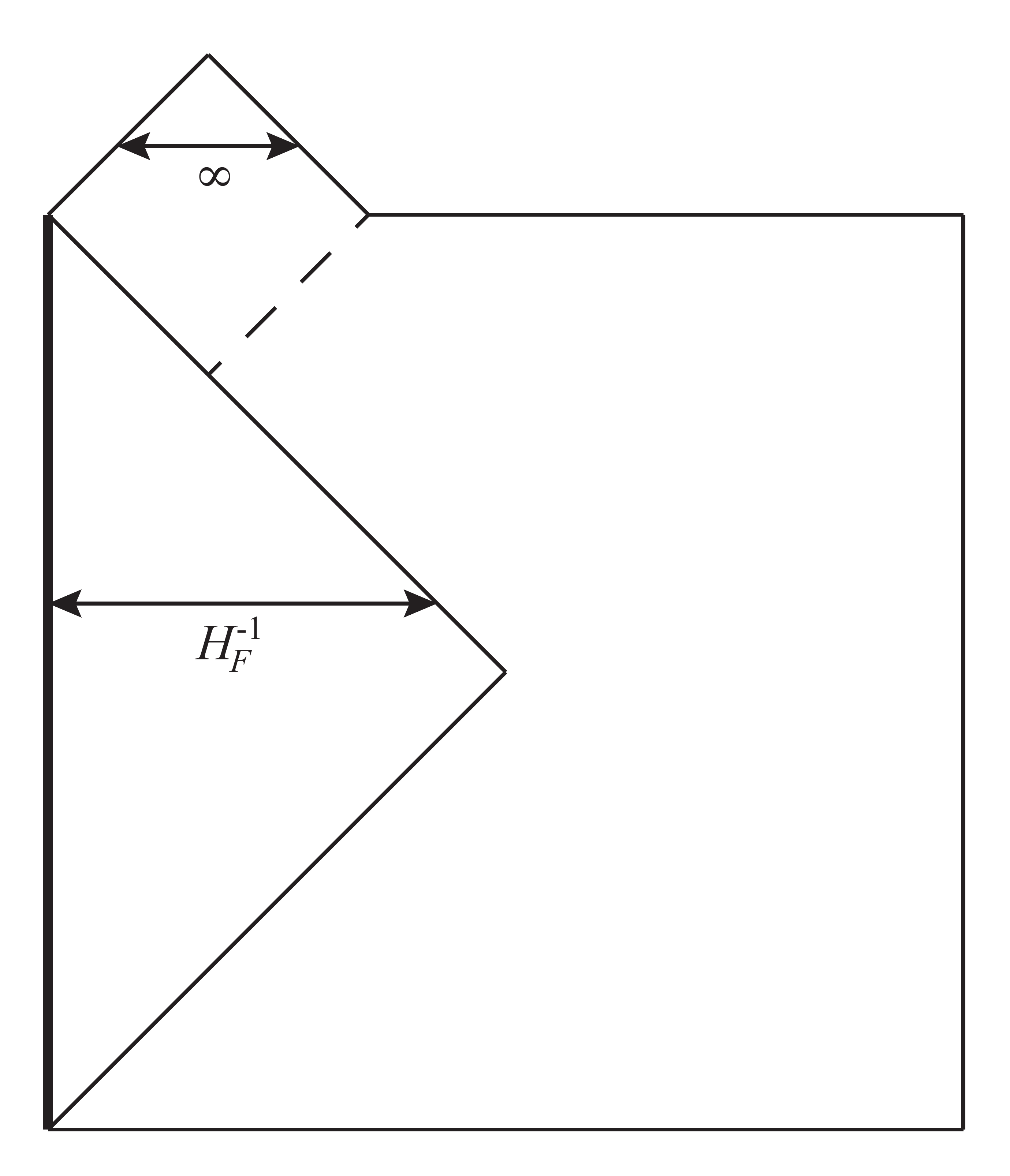}
  \end{center}
  \caption[Conformal diagrams for de~Sitter space with a false vacuum]{
    Conformal diagrams for de~Sitter space with a false vacuum.
    The first two diagrams show the effect of a bubble (dashed line) nucleating within the northern-hemisphere causal patch, leading to lower-energy de~Sitter [left] and Minkowski [middle], while the third shows a false-vacuum region that does not experience any bubbles [right].
    $H_F^{-1}$ and $H_T^{-1}$ are the Hubble radii of the false and true vacua, with the latter being infinite in the Minkowski case.
    In both cases the true horizon is larger than the Hubble radius in the false vacuum; in the left-hand diagram, it becomes equal to the horizon in the true vacuum, while on the right it becomes infinitely large.
    In either of these cases, excitations can leave the apparent horizon in the false vacuum while remaining inside the true horizon.
    On the right, the observer at the North pole remains in the false vacuum state forever, although there are bubbles outside their horizon.}
\label{conformal-fig2}
\end{figure}

Different cases of interest for bubble nucleation are shown in Figure~\ref{conformal-fig2}.
An observer at the north pole in the de~Sitter diagram could witness the nucleation of a bubble to a lower-energy de~Sitter vacuum, or to a Minkowski vacuum (the triangular ``hat''), or avoid seeing bubbles at all.
The probability of seeing a bubble along any specified geodesic asymptotes to 1, but for a sufficiently small nucleation rate, the physical volume of space remaining in the false vacuum grows with time.

Next we turn to uptunneling from one false-vacuum state to another of even higher energy.%
\footnote{We thank Stefan Leichenauer and Paul Steinhardt for discussions of these issues.}
In the true vacuum, we could straightforwardly argue that the spirit of the cosmic no-hair theorem is obeyed: excitations leave the horizon and the system approaches its lowest-energy eigenstate.
In the false vacuum, the argument is not so clean, since there are no true energy eigenstates to approach.
Nevertheless, the physical situation is quite similar.
The Hilbert space is still infinite-dimensional, so we do not expect recurrences, and excitations within a patch can readily leave the horizon, leaving us in the perturbative vacuum.
Once again, there exist instantons reflecting the overlap of the perturbative false vacuum with excited states.
Just as above, however, we cannot interpret these instantons as true dynamical processes; they signal, instead, that the actual state of the field in the false vacuum is not entirely concentrated around $\varphi_F$, but also has some support on field values corresponding to potential minima of higher energies.
We said above that excitations around the perturbative false vacuum state dissipate away, resulting in a state quiescent with respect to perturbative fluctuations.
Incorporation of instanton effects simply shifts the quiescent state slightly, like it did for the true vacuum state.
It is this quiescent state that is physical, not the perturbative vacua connected by instanton transitions.

We can see, in particular, that the wave function cannot start in a quiescent state and then split into two branches, describing uptunneling: uptunneling represents a decrease rather than an increase of entropy, so it can only be a Boltzmann fluctuation rather than a branching of the wave function.
The timescale over which the perturbative vacuum relaxes to the physical one (presumably with slightly smaller effective cosmological constant) will be governed by the barrier-penetration factor connecting the false vacuum to higher-energy minima.
That factor also governs the rate for uptunneling to such minima.
Therefore, we expect a relatively short window in which uptunneling can happen before the state relaxes, after which the rate of uptunneling falls to zero.
We can check this picture by considering the limit in which the barrier between false and true vacuum becomes infinitely large.
In that case, transitions from the false to true vacua are suppressed, and the behavior of the false vacuum should increasingly resemble our picture of the true vacuum.
This is precisely what we have found: in both cases, uptunneling is forbidden (more precisely, the rate of uptunneling is suppressed as excitations dissipate away).

While these results are not rigorous, they provide a strong indication that false-vacuum states in semiclassical quantum gravity either decay or asymptote to quiescent states that are free of dynamical fluctuations.

\subsection{Complementarity in a Landscape}
\label{falsevacuum-complementarity}

We now consider theories with multiple vacua, each labeled by a field expectation value $\varphi_i$, in the context of horizon complementarity.
In this case the Hilbert space appropriate to a single vacuum~\eqref{hilbertcomp} is promoted to a direct sum, with one term for each semiclassical patch geometry:
\begin{equation}
  \Hilbert = \bigoplus_i \Hilbert_{\textrm{bulk}}^{(i)}
  \otimes \Hilbert_{\textrm{boundary}}^{(i)} \ .
\end{equation}
The structure is similar to that of Fock space~\cite{Nomura:2011dt,Nomura:2011rb}.
The dimensionality of the entire Hilbert space is the sum of the dimensions of each term, $\dim\Hilbert^{(i)} = e^{2S_{\rm dS}^{(i)}} = \exp(6\pi\Lambda_i^{-1})$.
There are two cases of interest: the finite-dimensional case where every vacuum has $\Lambda>0$ and the infinite-dimensional case where there is at least one vacuum with $\Lambda=0$.
(As mentioned previously, we do not consider vacua with $\Lambda <0$, as transitions into them lead to singularities.)

If all vacua have $\Lambda >0$, the situation is very similar to the single-vacuum case discussed in Subsection~\ref{singledscomplementarity}.
Exact energy eigenstates, including the lowest-energy vacuum state, will be stationary and no dynamical fluctuations will occur.
The vacuum will feature a de~Sitter semiclassical geometry with the field value concentrated near the true minimum, although it will not be a field eigenstate.
Generic states, however, will not be stationary, and in a finite-dimensional Hilbert space there is no room for excitations to dissipate outside the horizon, so recurrences are expected.

Now consider theories with at least one vacuum having $\Lambda = 0$, as might be expected in supersymmetric or string theories.
The future development of the spacetime includes census-taker observers living in a Minkowski ``hat''~\cite{Susskind:2007pv,Sekino:2009kv}, as shown in the middle and right diagrams of Figure \ref{conformal-fig2}.
The Hilbert space of the full theory is then infinite dimensional, and such observers have access, in principle, to an infinite amount of information.

From \eqref{eq:zurek_fluctuations}, the rate of Boltzmann fluctuations goes to zero (and the timescale for recurrences goes to infinity) for infinite-dimensional Hilbert spaces, where $\Lambda_T = 0$.
Of course, there are no dynamical fluctuations in the true Minkowski vacuum.
But we can make a stronger statement: the rate of the fluctuations will asymptote to zero even in the false vacua.
The intuition is that states with excitations around false-vacuum geometries are more likely to decay than the vacuum states themselves.
So time evolution will skew the population of false vacua towards states that are stationary except for the possibility of decay by bubble nucleation, \textit{i.e.}~quiescent in the sense of the previous subsection.
After a high-energy vacuum decays to a lower-energy one, transient excitations will allow for the existence of Boltzmann fluctuations, but the excited states will again preferentially decay.
The surviving configurations will become effectively stationary, and the Boltzmann fluctuation rate will asymptote to zero, rather than to a nonzero constant.
We therefore expect only a finite (and presumably small) number of Boltzmann fluctuations in a landscape of vacua that includes a Minkowski vacuum.

This intuition can be bolstered by an analogy to one-dimensional quantum mechanics in the presence of a barrier.
Consider once again a particle of mass $m$ and energy $E$ moving in a potential $V(x)$ schematically similar to the false-vacuum potential shown in Figure~\ref{vacua-fig}.
The particle can escape the well by tunneling through the barrier.
A wave packet initially in the potential well will leak out, and the WKB approximation relates the wave functions on either side of the potential:
\begin{equation}
  \frac{\psi(x_{e})}{\psi(x_{0})}=\exp\left(-\frac{1}{\hbar}
  \int_{x_{0}(E)}^{x_{e}(E)}\sqrt{2m(V(x)-E)}dx\right)\equiv e^{-\gamma/2} \ ,
\end{equation}
where $x_0(E)$ and $x_e(E)$ are the starting and ending points for the region where the particle ``has negative energy,'' so $V(x_0(E))=V(x_e(E))=E$.
The escape probability is simply $e^{-\gamma}$, and the tunneling rate is given by the product of this probability with some characteristic frequency:
\begin{equation}
  R=f(E)e^{-\gamma} \ .
\end{equation}
The classic barrier penetration problem considers a square-well potential, in which the bound particle has a position-independent momentum, $p(E)=\sqrt{2m(E-V)}$, and a characteristic ``collision frequency'', $f(E)=p(E)/(2mx_{0})$.
Here, we assume a more general potential, so the momentum is a function of both $E$ and $x$, and the frequency will be given by some integral over positions inside the well.
The exact expression is not important for us---we assume only that the frequency is an increasing function of $E$,  $f^\prime (E)>0$.
Then, the energy dependence of the tunneling rate is
\begin{equation}
  \frac{dR}{dE}=f^\prime(E)e^{-\gamma}-\frac{2}{\hbar}f(E)e^{-\gamma}
  \int_{x_{0}(E)}^{x_{e}(E)}\left[-\frac{2}{\sqrt{2m(V(x)-E)}} \right] \ ,
\end{equation}
which is manifestly positive.
(We have used the fact that $V(x_0)-E=V(x_e)-E=0$ to eliminate the terms which arise from varying the limits of integration.)

This simple exercise demonstrates an intuitively sensible result: among states trapped behind a barrier, those with higher energy tunnel out more quickly.
In the case of the cosmological false vacuum, the analogous statement is that excited states of the perturbative Hamiltonian undergo false-vacuum decay more rapidly.

In complementarity, we see that only in the case of a Minkowski true vacuum can recurrences and Boltzmann fluctuations be avoided entirely.
A version of this phenomenon---the crucial difference in the long-term quantum evolution of landscapes with and without $\Lambda=0$ vacua---has been previously noted in a slightly different context~\cite{Bousso:2011up,Nomura:2011rb}.
There, it was pointed out that quantum measurements in a false-vacuum state will decohere by becoming entangled with environment degrees of freedom, but they must eventually recohere if the total Hilbert space is finite-dimensional.
In infinite-dimensional Hilbert spaces, in contrast, decoherence can persist forever.
This argument is analogous to our own, in that such models are largely free of Boltzmann fluctuations.

\section{Consequences}
\label{consequences}

\subsection{Boltzmann Brains}

In the conventional picture, because de~Sitter space has a temperature, it experiences thermal fluctuations that lower the entropy by $\Delta S$ with a finite rate proportional to $e^{-\Delta S}$.
If the Hartle-Hawking vacuum is eternal, then all dynamical fluctuations that fit within a horizon volume are produced an infinite number of times inside each such volume.
Such fluctuations could contain conscious observers like ourselves~\cite{Dyson:2002pf,Albrecht:2004ke,Bousso:2006xc,Page:2005ur,Page:2006dt,Page:2006hr,Page:2006ys,Page:2006nt,Page:2009mc}.
Due to the exponential suppression of lower-entropy states, the fluctuations containing observers---even the ones that contain exact copies of our own brains---that occurred most frequently would look entirely unlike the world we observe.
In particular, fluctuations containing the room you are reading this paper in would be vastly more likely than fluctuations containing all of Earth, let alone the entire observable universe, and the momentary coalescence of your brain thinking the precise thoughts you are having right now out of thermal equilibrium would be likelier still.
If this conclusion were correct, we would not be able to trust our memories or our (supposed) observations, a solution inconducive to the practice of science.

We have argued, however, that this situation is less generic in de~Sitter cosmologies than is often supposed.%
\footnote{For related work that questions the validity of Boltzmann brains for decoherence-based reasons, see~\cite{Davenport:2010jy,Gott:2008ii,Aaronson:2013ema}.
For the need for Hilbert space to be infinite-dimensional, see~\cite{Carroll:2008yd}.}
The appearance of Boltzmann brains is avoided in the context of QFT in eternal de~Sitter space or in a landscape with a terminal Minkowski vacuum (with or without complementarity).
In these cases, the dimension of the Hilbert space is infinite, so the recurrence time also goes to infinity, and the (possibly false) de~Sitter vacuum becomes quiescent.
If the horizon volume is initially in an excited state (as it is if the dark energy is a positive cosmological constant), then the cosmic no-hair theorem dictates that correlations fall off exponentially with time as the excitations leave the horizon.
The total number of Boltzmann brains will thus be finite and presumably small, given the vast exponential suppression of macroscopic fluctuations.
Thus, if enough observers are produced before de~Sitter space approaches the vacuum (\textit{e.g.}, in a period of structure formation) the vast majority of observers can, in fact, trust their memories and observations.
This conclusion opens the door for many multiverse models that might have been discounted because of a Boltzmann brain problem, and could help resolve potential tensions with low-energy physics~\cite{Boddy:2013qma}.

\subsection{Landscape Eternal Inflation}
\label{landscape}

Another kind of fluctuation into a lower-entropy state that is often invoked in de~Sitter cosmology is uptunneling from one de~Sitter vacuum state to another one of higher energy~\cite{Lee:1987qc,Aguirre:2011ac}.
Processes such as this can be crucial for populating an entire landscape of vacua, starting from a state concentrated on any particular field value.

Uptunneling is conceptually very similar to the standard picture of a fluctuation into a Boltzmann brain: a vacuum in a thermal state undergoes a transition to a lower-entropy configuration with probability $e^{-\Delta S}$.
The situation is the time-reverse of the well-known process of vacuum decay, which results in the production of particles and an increase in entropy.
The analysis presented in this paper leads to an analogous conclusion to that of the last subsection: if the total Hilbert space is infinite-dimensional, excitations around any particular false vacuum will dissipate.
As discussed in Section~\ref{falsevacuum}, the system will relax to a (perturbative, semi-perturbative, or true) vacuum state, not a state of definite field value.
The state becomes quiescent, and the rate of Boltzmann fluctuations asymptotes to zero.

Note that eternal inflation is still conceivable: uptunneling is suppressed, but downtunneling proceeds as usual, and different branches of the wave function will correspond to different distributions of bubbles in a semiclassical spacetime background.
If the field starts out in a metastable vacuum, then the portion of it that remains there (on any one branch) is rewarded with greater volume production.
Almost every world line will intersect a bubble of lower-energy vacuum, but if the tunneling rate is low enough to avoid percolation, the physical volume remaining in the high-energy vacuum grows without bound, as depicted in the rightmost diagram in Figure~\ref{conformal-fig2}.
In this sense inflation continues forever.

On the other hand, it is clear that the details of eternal inflation in a landscape of vacua will change.
In particular, the conclusions of the previous section suggest a reinterpretation of the rate equations for eternal inflation that relate the probabilities of transitions between different vacua~\cite{Garriga:1997ef, Linde:2006nw, Nomura:2011dt}.
Consider the simple landscape of Figure~\ref{vacua-fig}, with minima located at field values $\varphi_F$ and $\varphi_T$, respectively.
In the standard presentation, \textit{e.g.}~\cite{Garriga:1997ef}, the rate equations for a two-minimum landscape read
\begin{equation}
  \frac{dp_{f}}{d\tau}=-\kappa_{f}p_{f}+\kappa_{t}p_{t} \ , \qquad
  \frac{dp_{t}}{d\tau}=-\kappa_{t}p_{t}+\kappa_{f}p_{f} \ ,
\end{equation}
where $\kappa_f$ and $\kappa_t$ are transition probabilities per unit proper time.
The usual interpretation is that $\kappa_f$ ($\kappa_t$) represents the probability to transition from the false (true) vacuum to the true (false) one.
But we have argued that, in the long-time limit, the probability to transition from the true to a false vacuum falls to zero.
However, both the true and the false vacuum states have nonzero overlap with the states of any definite field value, so heuristically we may think of the true vacuum, for example, as containing an exponentially small piece with field value near $\varphi_F$.
The rate equations should essentially be interpreted as probabilities to transition between states of definite field value in an (unrealistic) idealization where an observer is measuring the value of the field at regular intervals.
In the real universe, where there is no external observer and the wave function evolves unitarily, the state simply evolves toward the true vacuum as time passes.
Dynamical fluctuations in de~Sitter space do not provide a mechanism for populating an entire landscape with actual semiclassical geometries centered on different vacua and living on different branches of the wave function.

With horizon complementarity, this picture changes somewhat.
If the true vacuum is de~Sitter, Hilbert space is finite-dimensional, and Boltzmann fluctuations will lead to true transitions between states concentrated at different minima of the potential.
If the true vacuum is Minkowski, on the other hand, Hilbert space is infinite-dimensional, and the above discussion is once again valid.

\subsection{Inflationary Perturbations}

The absence of dynamical fluctuations in the de~Sitter vacuum might seem to call into question the standard picture of the origin of density perturbations in inflation.
In this case, however, the conventional wisdom gets the right answer; our approach leaves the standard predictions for density and tensor fluctuations from inflation essentially unaltered.
The basic point is that the quantum state of light fields can remain coherent during inflation itself, and possess (non-dynamical) vacuum fluctuations, but then experience decoherence and branching of the wave function when entropy is generated at reheating.

We can describe the Hilbert space during inflation as a product of the quantum states of the large-scale homogeneous background $\varphi(t)$ (macroscopic) perturbations and the small-scale (microscopic) perturbations:
\begin{equation}
  \Hilbert = \Hilbert_{\varphi(t)} \otimes \Hilbert_{\rm macro}
  \otimes \Hilbert_{\rm micro} \ .
\end{equation}
The small-scale perturbations, including the specific microstates of individual photons and other particles, are unobservable, in the same way that individual atoms and molecules are unobservable in an ordinary box of gas.
They serve as an environment we can trace over to understand the state of the observable large-scale perturbations.
During inflation, the overall quantum state approaches a factorizable form, as excitations dissipate and perturbations approach their lowest-energy states:
\begin{equation}
  \ket{\Psi_{\rm inflation}} = \ket{\varphi(t)} \otimes \ket{0}_{\rm macro}
  \otimes \ket{0}_{\rm micro} \ .
\end{equation}
The state $\ket{0}_{\rm macro}$ has a nonzero variance for the field operator $\varphi$, as calculated in standard treatments, but its quantum coherence is maintained.%
\footnote{One might imagine that decoherence occurs because modes become super-Hubble-sized, and we should trace over degrees of freedom outside the horizon.
This reasoning is not quite right, as such modes could (and often do) later re-enter the observable universe; they become larger than the Hubble radius during inflation but never leave the true horizon.}

At reheating, however, entropy is generated.
Energy in the inflaton is converted into a dense, hot plasma with many degrees of freedom.
The specific form of the microscopic perturbations will depend on the state of the macroscopic perturbations; these factors become entangled, producing a state of the form
\begin{equation}
  \ket{\Psi_{\rm reheating}} = \ket{\varphi(t)} \otimes
  \left[\sum_i \ket{\delta\varphi_i}_{\rm macro} \otimes
    \left(\sum_\mu \ket{\delta\varphi_{i, \mu}}_{\rm micro} \right)\right] \ .
\end{equation}
Tracing over the microscopic fluctuations leaves a mixed-state density matrix for the macroscopic fluctuations, inducing decoherence~\cite{Polarski:1995jg,Lombardo:2005iz,Martineau:2006ki,Burgess:2006jn,Kiefer:2006je,Prokopec:2006fc}.
By this process, the unique quantum state of the inflaton field evolves into a large number of decohered branches, each with a specific pattern of perturbations such as we observe in the CMB, with statistics given by the Born rule.
In effect, reheating acts as an explicit measurement process.
We, therefore, expect that the standard calculations of scalar and tensor fluctuations in any given inflationary model are unaffected by the considerations in this paper.

\subsection{Stochastic Eternal Inflation}

We next turn to the possibility of eternal inflation in a slow-roll potential, as distinguished from a landscape of false vacua.
The traditional approach to this scenario makes use of the stochastic approximation, which treats the inflaton field value in the slow-roll regime as a stochastic variable, undergoing a random walk~\cite{Vilenkin:1983xq,1987ZhETF..92.1137G,Goncharov:1987ir}; for recent treatments see~\cite{Creminelli:2008es,Dubovsky:2011uy,Martinec:2014uva}.
Consider the case of a power-law potential,
\begin{equation}
  V(\varphi) = \frac{\lambda\varphi^{2n}}{2n \mpl^{2n-4}} \ .
\end{equation}
In a single Hubble time, the expectation value of the field decreases by
\begin{equation}
  \Delta\varphi = \frac{n\mpl^{2}}{4\pi\varphi} \ ,
\end{equation}
but the dispersion of perturbations around this value is
\begin{equation}
  \Delta^{2}=\avg{\delta\varphi^{2}} = \frac{H^{3}}{4\pi^{2}}t \ .
\end{equation}
In a Hubble time $H^{-1}$, we have $\Delta=H/2\pi$.

Now comes the critical step. In the stochastic approximation, one asserts that $\Delta$ represents an RMS fluctuation amplitude
\begin{equation}
  \left|\delta\varphi\right| = \frac{H}{2\pi} \ ,
\end{equation}
and that the effective value of the inflaton in a given Hubble patch should be treated as a random variable drawn from a distribution with this amplitude.
Above a critical field value,
\begin{equation}
  \varphi^{*} = \lambda^{-1/(2n+2)} \mpl \ ,
\end{equation}
the fluctuations dominate, $\left|\delta\varphi\right|\gg\Delta\varphi$.
In this picture, to an excellent approximation, $\varphi$ undergoes a random walk with time step $H^{-1}$ and step size $\left|\delta\varphi\right|$.
Causality dictates that each horizon area undergoes these fluctuations independently.
Every Hubble time, when a horizon volume grows by a factor $e^3\sim20$, the field value in approximately 10 of the new horizon volumes is larger than its parent.
In fact, this statement is a much stronger condition then required for eternal inflation.
It suffices for only one of these volumes to move upward on the potential: $\left|\delta\varphi\right|\approx\mathcal{O}\left(\Delta\varphi/20\right)$.

The stochastic-approximation approach to eternal inflation is in tension with the analysis presented in this paper.
As we have argued in Section~\ref{fluctuations}, quantum fluctuations in closed systems near equilibrium cannot be treated as classical random variables.
Fluctuations $\delta\varphi$ only become real when they evolve into different decoherent branches of the wave function and generate entropy (what we have called measurement-induced fluctuations).
For the perturbations we observe in the CMB, this entropy source is provided by reheating.
But precisely in the slow-roll regime, where the stochastic inflation story is invoked, there is no entropy production, no measurement or decoherence, and no branching of the wave function.
All that happens during a Hubble time is a decrease in the classical field expectation value, $\Delta\varphi$.
There is no quantum-dominated regime; the field simply rolls down its potential.

\begin{figure}[t]
  \begin{center}
    \includegraphics[width=0.7\textwidth]{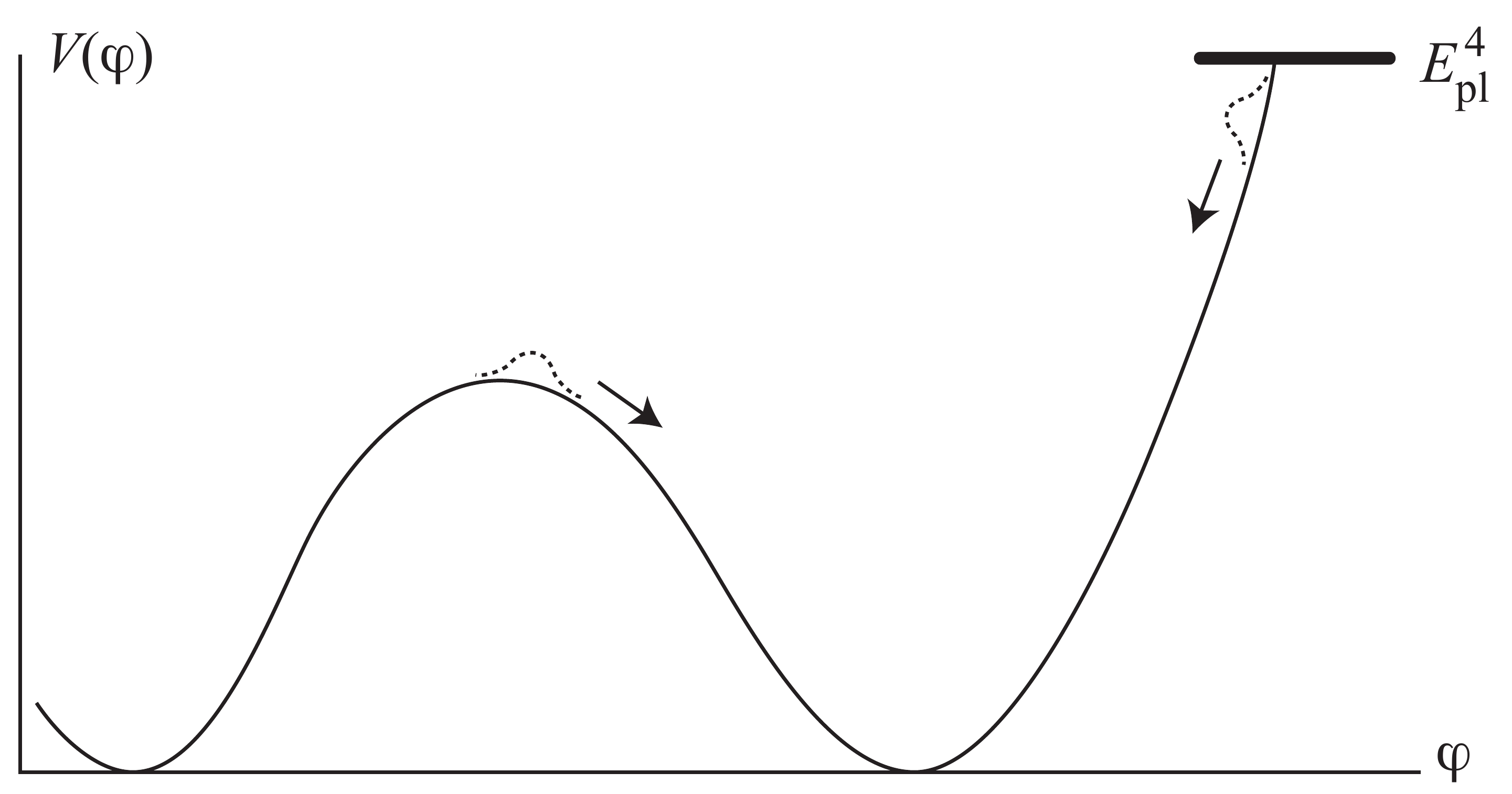}
  \end{center}
  \caption[Potential supporting different kinds of inflation]{
    Potential supporting different kinds of inflation.
    Dashed lines are schematic representations of two different initial quantum states for the field.
    If the field begins at the right edge near the Planck cutoff, we expect it to evolve smoothly to the non-inflationary regime at the bottom of the potential.
    In contrast, if it begins at the top of a hill, it is plausible to imagine that part of the wave function remains in an inflationary state for arbitrarily long periods of time (although the amplitude for that branch of the wave function will be monotonically decreasing).}
\label{inflation-fig}
\end{figure}

A more honest approach to eternal inflation would be to take the quantum nature of the dynamics seriously, and investigate the evolution of the wave function describing the coupled background and perturbations; we hope to study this more carefully in future work.
Nevertheless, it is possible to draw some qualitative conclusions by considering the evolution of a wave packet in field space representing the homogeneous mode.
If the initial state of the field has support near a local maximum of the potential, inflation is plausibly eternal: part of the wave packet will roll down the potential, eventually couple to perturbation modes, and experience decoherence, while part will remain near the maximum and continue to inflate.
In contrast, if the field is slowly rolling down a monotonic portion of the potential---as expected for a polynomial potential with a Planck-scale energy density cutoff---it will reach the bottom of the potential, and the inflationary phase will end in a finite time and after a finite number of $e$-folds.
These two possibilities are portrayed in Figure~\ref{inflation-fig}.
We note that the simplest inflaton potentials, monomial power-laws $V(\varphi)\sim\varphi^n$, do not have saddle points and should thus avoid eternal inflation given a Planck-scale cutoff.
(The recent BICEP2 detection~\cite{Ade:2014xna} of large-scale B-mode polarization in the CMB, if interpreted as a tensor/scalar ratio $r\sim0.2$, is well fit by an $m^2\varphi^2$ potential.)

For a field on the monotonic portion of the potential, one might object that, even once the field has rolled down, some portion of the wave function will always remain arbitrarily close to the maximum allowed value of the potential, \textit{e.g.} the Planck-scale cutoff, just as a wave packet is supported throughout all of space despite being concentrated around a single point.
This reasoning is correct, but it does not imply that there are some portions of the wave function where the end of inflation is postponed.
The problem with this interpretation was already noted in Subsections~\ref{semiclassical} and \ref{landscape}: states of definite field value are not the same as states of definite energy density.
In the slow-roll approximation, the cosmic no-hair theorem acts to bring the inflaton field to the appropriate vacuum state---a state of energy density corresponding perturbatively to de Sitter space with the appropriate cosmological constant.
Each such state has nonzero overlap with the states of definite field value, but the cosmic no-hair theorem guarantees (to the extent that the slow-roll approximation is valid, so that no entropy is produced) that the field is driven into the appropriate false vacuum state, and then rolls smoothly to states with lower and lower energy density until the point that inflation ends.
Again, there is negligible entropy production, no measurement-induced fluctuations, and no branching during this period---the inflaton remains in a single coherent state until reheating occurs.

To gain intuition for the points we make above, it is useful to consider applying the stochastic approximation to a free massive scalar field in eternal de~Sitter space itself.
At the minimum of the potential $V=m^2\varphi^2/2$, it is clear that the classical change $\Delta\varphi$ vanishes while the quantum variance $\delta\varphi$ does not, so the system is automatically in the quantum-dominated regime.
If the stochastic approximation is applied, we expect occasional fluctuations of the field to very large values, leading to rapid inflation in those regions but not in others.
In other words, if the stochastic picture is valid, one is led to the conclusion that de~Sitter space with a massive scalar field has a runaway instability, in contrast with the usual view that there is a lowest-energy eigenstate with a stable semiclassical geometry (\textit{c.f.}~\cite{Bousso:1998bn}).
In light of the above, we interpret this purported instability differently: it indicates a problem with the stochastic approximation, not with de~Sitter space itself.
The vacuum state of the scalar field is not a state of definite field value, although it is centered around the minimum of the potential.
Rather, the state has overlap with all field values, at least up to a potential Planck-scale energy density cutoff.
But we do not interpret the de~Sitter vacuum as an unstable superposition of different field values expanding at different rates.
Instead, we say that the field is in a single state, the vacuum, with a definite energy density given by the cosmological constant $\Lambda$.

\subsection{Other Formulations of Quantum Mechanics}

Throughout this paper we have worked in the context of the Everett/Many-Worlds formulation of quantum mechanics, in which a single wave function evolves unitarily in Hilbert space according to the Schr\"odinger equation.
Our conclusions could be dramatically altered in other formulations.
We will not explore these possibilities in detail here, but we briefly mention two alternatives.

One would be the de~Broglie-Bohm approach, in which the variables include both a wave function and variables in a separate configuration space~\cite{Bohm:1951xw,Bohm:1951xx,Durr:2003gu,Struyve:2007fs}.
In such a theory, the wave function could be completely stationary (as in the de~Sitter vacuum), but the configuration-space variables could still fluctuate.
What we think of as a stationary thermal state in the Everett approach would be more closely analogous to a thermal distribution function in classical statistical mechanics; while the density operator is stationary, the underlying state could still be evolving in time.
We might, therefore, observe dynamical fluctuations even in equilibrium.
Recent work has argued, however, that in practice these Boltzmann fluctuations can be avoided in Bohmian cosmology \cite{Goldstein:2015mha}.

Another alternative is a stochastic dynamical-collapse theory, such as the Ghirardi-Rimini-Weber (GRW) model~\cite{grw85,Ghirardi:1985mt}.
Set in the context of nonrelativistic, many-particle quantum mechanics, the wave function has a fixed probability per particle per unit time of spontaneously collapsing to a localized position.
Entanglement between particles induces an effective, ongoing ``measurement'' of macroscopic systems.
There is not a well-developed GRW model for QFT in de~Sitter space, but the philosophy of the approach leads us to expect that a thermal state would experience true fluctuations; the possibility of dynamical collapse introduces a new kind of time-dependence that would be absent in equilibrium in Everettian quantum theory.
It seems we are dealing with one of the rare cases in which one's favorite formulation of quantum mechanics can drastically affect one's expectation for how observable quantities evolve.

\section{Conclusions}

Quantum variables are not equivalent to classical stochastic variables.
They can be related by the appearance of measurement-induced fluctuations, which require entropy generation, decoherence, and branching of the wave function.
In stationary states, entropy is not generated, and the wave function remains fixed; therefore, there are no dynamical fluctuations, and treating a quantum field as a classical stochastic field is inappropriate.
We have argued that this shift in thinking has important consequences for the cosmology of de~Sitter space, since de~Sitter regions tend to approach a stationary thermal state.
In particular, if the true Hilbert space is infinite-dimensional (as is the case in QFT in curved spacetime or in horizon complementarity in the presence of a Minkowski vacuum), de~Sitter vacua settle down and do not fluctuate.
There are no Boltzmann brains in such states, relieving a major problem for many multiverse cosmological models.
On the other hand, we also suggest there is neither uptunneling to higher-energy vacua nor stochastic fluctuations up a slow-roll potential, implying that eternal inflation is much less generic than often supposed.
A better understanding of complementarity and the correct formulation of quantum mechanics will help establish what happens in the real universe.

\section*{Acknowledgments}

We have benefited from helpful discussions with Scott Aaronson, Charles Bennett, Alan Guth, James Hartle, Stefan Leichenauer, Spyridon Michalakis, Don Page, John Preskill, Jess Riedel, Charles Sebens, Paul Steinhardt, and several participants at the Foundational Questions Institute conference on The Physics of Information (though they might not agree with our conclusions).
This research is funded in part by DOE grant DE-FG02-92ER40701 and by the Gordon and Betty Moore Foundation through Grant 776 to the Caltech Moore Center for Theoretical Cosmology and Physics.

\bibliography{brains-bib}

\providecommand{\href}[2]{#2}\begingroup\raggedright\begin{thebibliography}{100}

\bibitem{Bunch:1978yq}
T.~Bunch and P.~Davies, ``{Quantum Field Theory in de Sitter Space:
  Renormalization by Point Splitting},'' {\em Proc. Roy. Soc. Lond.} {\bfseries
  A360} (1978) 117.

\bibitem{Bunch:1978yw}
T.~Bunch and P.~Davies, ``{Nonconformal Renormalized Stress Tensors in
  Robertson-Walker Space-Times},''
  \href{http://dx.doi.org/10.1088/0305-4470/11/7/018}{{\em J. Phys.} {\bfseries
  A11} (1978) 1315}.

\bibitem{Hartle:1976tp}
J.~Hartle and S.~Hawking, ``{Path Integral Derivation of Black Hole
  Radiance},'' \href{http://dx.doi.org/10.1103/PhysRevD.13.2188}{{\em Phys.
  Rev.} {\bfseries D13} (1976) 2188}.

\bibitem{Gibbons:1977mu}
G.~Gibbons and S.~Hawking, ``Cosmological event horizons, thermodynamics, and
  particle creation,'' \href{http://dx.doi.org/10.1103/PhysRevD.15.2738}{{\em
  Phys. Rev.} {\bfseries D15} (1977) 2738}.

\bibitem{Banks:2000fe}
T.~Banks, ``{Cosmological breaking of supersymmetry? or Little lambda goes back
  to the future 2},'' \href{http://dx.doi.org/10.1142/S0217751X01003998}{{\em
  Int. J. Mod. Phys.} {\bfseries A16} (2001) 910},
  \href{http://arxiv.org/abs/hep-th/0007146}{{\ttfamily arXiv:hep-th/0007146
  [hep-th]}}.

\bibitem{Banks:2001yp}
T.~Banks and W.~Fischler, ``{M theory observables for cosmological
  space-times},'' \href{http://arxiv.org/abs/hep-th/0102077}{{\ttfamily
  arXiv:hep-th/0102077 [hep-th]}}.

\bibitem{Lyth:2009zz}
D.~H. Lyth and A.~R. Liddle, {\em {The Primordial Density Perturbation:
  Cosmology, Inflation and the Origin of Structure}}.
\newblock Cambridge University Press, Revised edition 2009.

\bibitem{Dodelson:2003ft}
S.~Dodelson, {\em Modern Cosmology}.
\newblock Academic Press, San Diego, CA, 2003.

\bibitem{Baumann:2009ds}
D.~Baumann, ``{TASI Lectures on Inflation},''
  \href{http://arxiv.org/abs/0907.5424}{{\ttfamily arXiv:0907.5424 [hep-th]}}.

\bibitem{Vilenkin:1983xq}
A.~Vilenkin, ``{The Birth of Inflationary Universes},''
\href{http://dx.doi.org/10.1103/PhysRevD.27.2848}{{\em Phys. Rev.} {\bfseries
  D27} (1983) 2848}.

\bibitem{1987ZhETF..92.1137G}
A.~S. {Goncharov} and A.~D. {Linde}, ``{Global structure of the inflationary
  universe},'' {\em Zh. Eksp. Teor. Fiz.} {\bfseries 92} (1987) 1137.

\bibitem{Goncharov:1987ir}
A.~Goncharov, A.~D. Linde, and V.~F. Mukhanov, ``{The Global Structure of the
  Inflationary Universe},''
\href{http://dx.doi.org/10.1142/S0217751X87000211}{{\em Int. J. Mod. Phys.}
  {\bfseries A2} (1987) 561}.

\bibitem{Weinberg:1987dv}
S.~Weinberg, ``{Anthropic Bound on the Cosmological Constant},''
  \href{http://dx.doi.org/10.1103/PhysRevLett.59.2607}{{\em Phys. Rev. Lett.}
  {\bfseries 59} (1987) 2607}.

\bibitem{Bousso:2000xa}
R.~Bousso and J.~Polchinski, ``{Quantization of four form fluxes and dynamical
  neutralization of the cosmological constant},''
  \href{http://dx.doi.org/10.1088/1126-6708/2000/06/006}{{\em JHEP} {\bfseries
  06} (2000) 006}, \href{http://arxiv.org/abs/hep-th/0004134}{{\ttfamily
  arXiv:hep-th/0004134 [hep-th]}}.

\bibitem{Kachru:2003aw}
S.~Kachru, R.~Kallosh, A.~D. Linde, and S.~P. Trivedi, ``{De Sitter vacua in
  string theory},'' \href{http://dx.doi.org/10.1103/PhysRevD.68.046005}{{\em
  Phys. Rev.} {\bfseries D68} (2003) 046005},
  \href{http://arxiv.org/abs/hep-th/0301240}{{\ttfamily arXiv:hep-th/0301240
  [hep-th]}}.

\bibitem{Susskind:2003kw}
L.~Susskind, ``{The Anthropic landscape of string theory},'' in {\em The Davis
  Meeting On Cosmic Inflation}.
\newblock Mar., 2003.
\newblock \href{http://arxiv.org/abs/hep-th/0302219}{{\ttfamily
  arXiv:hep-th/0302219 [hep-th]}}.

\bibitem{Denef:2004ze}
F.~Denef and M.~R. Douglas, ``{Distributions of flux vacua},''
  \href{http://dx.doi.org/10.1088/1126-6708/2004/05/072}{{\em JHEP} {\bfseries
  05} (2004) 072}, \href{http://arxiv.org/abs/hep-th/0404116}{{\ttfamily
  arXiv:hep-th/0404116 [hep-th]}}.

\bibitem{Lee:1987qc}
K.-M. Lee and E.~J. Weinberg, ``{Decay of the True Vacuum in Curved
  Space-time},''
\href{http://dx.doi.org/10.1103/PhysRevD.36.1088}{{\em Phys.~Rev.} {\bfseries
  D36} (1987) 1088}.

\bibitem{Aguirre:2011ac}
A.~Aguirre, S.~M. Carroll, and M.~C. Johnson, ``{Out of equilibrium:
  understanding cosmological evolution to lower-entropy states},''
  \href{http://dx.doi.org/10.1088/1475-7516/2012/02/024}{{\em JCAP} {\bfseries
  02} (2012) 024},
\href{http://arxiv.org/abs/1108.0417}{{\ttfamily arXiv:1108.0417 [hep-th]}}.

\bibitem{Dyson:2002pf}
L.~Dyson, M.~Kleban, and L.~Susskind, ``Disturbing implications of a
  cosmological constant,''
  \href{http://dx.doi.org/10.1088/1126-6708/2002/10/011}{{\em JHEP} {\bfseries
  10} (2002) 011}.

\bibitem{Albrecht:2004ke}
A.~Albrecht and L.~Sorbo, ``{Can the universe afford inflation?},''
  \href{http://dx.doi.org/10.1103/PhysRevD.70.063528}{{\em Phys. Rev.}
  {\bfseries D70} (2004) 063528},
  \href{http://arxiv.org/abs/hep-th/0405270}{{\ttfamily arXiv:hep-th/0405270
  [hep-th]}}.

\bibitem{Bousso:2006xc}
R.~Bousso and B.~Freivogel, ``{A Paradox in the global description of the
  multiverse},'' \href{http://dx.doi.org/10.1088/1126-6708/2007/06/018}{{\em
  JHEP} {\bfseries 06} (2007) 018},
  \href{http://arxiv.org/abs/hep-th/0610132}{{\ttfamily arXiv:hep-th/0610132
  [hep-th]}}.

\bibitem{Banks:2002wr}
T.~Banks, W.~Fischler, and S.~Paban, ``{Recurrent nightmares? Measurement
  theory in de Sitter space},'' {\em JHEP} {\bfseries 12} (2002) 062,
\href{http://arxiv.org/abs/hep-th/0210160}{{\ttfamily arXiv:hep-th/0210160
  [hep-th]}}.

\bibitem{Spradlin:2001pw}
M.~Spradlin, A.~Strominger, and A.~Volovich, ``{Les Houches lectures on de
  Sitter space},''
\href{http://arxiv.org/abs/hep-th/0110007}{{\ttfamily arXiv:hep-th/0110007
  [hep-th]}}.

\bibitem{Wald:1983ky}
R.~M. Wald, ``Asymptotic behavior of homogeneous cosmological models in the
  presence of a positive cosmological constant,'' {\em Phys. Rev.} {\bfseries
  D28} (1983) 2118.

\bibitem{Hollands:2010pr}
S.~Hollands, ``{Correlators, Feynman diagrams, and quantum no-hair in deSitter
  spacetime},'' \href{http://dx.doi.org/10.1007/s00220-012-1653-2}{{\em Commun.
  Math. Phys.} {\bfseries 319} (2013) 1},
  \href{http://arxiv.org/abs/1010.5367}{{\ttfamily arXiv:1010.5367 [gr-qc]}}.

\bibitem{Marolf:2010nz}
D.~Marolf and I.~A. Morrison, ``{The IR stability of de Sitter QFT: results at
  all orders},'' \href{http://dx.doi.org/10.1103/PhysRevD.84.044040}{{\em Phys.
  Rev.} {\bfseries D84} (2011) 044040},
  \href{http://arxiv.org/abs/1010.5327}{{\ttfamily arXiv:1010.5327 [gr-qc]}}.

\bibitem{Stephens:1993an}
C.~R. Stephens, G.~'t~Hooft, and B.~F. Whiting, ``{Black hole evaporation
  without information loss},''
  \href{http://dx.doi.org/10.1088/0264-9381/11/3/014}{{\em Class. Quant. Grav.}
  {\bfseries 11} (1994) 621},
  \href{http://arxiv.org/abs/gr-qc/9310006}{{\ttfamily arXiv:gr-qc/9310006
  [gr-qc]}}.

\bibitem{Susskind:1993if}
L.~Susskind, L.~Thorlacius, and J.~Uglum, ``{The Stretched horizon and black
  hole complementarity},''
  \href{http://dx.doi.org/10.1103/PhysRevD.48.3743}{{\em Phys. Rev.} {\bfseries
  D48} (1993) 3743}, \href{http://arxiv.org/abs/hep-th/9306069}{{\ttfamily
  arXiv:hep-th/9306069 [hep-th]}}.

\bibitem{Parikh:2002py}
M.~K. Parikh, I.~Savonije, and E.~P. Verlinde, ``{Elliptic de Sitter space:
  dS/Z(2)},'' \href{http://dx.doi.org/10.1103/PhysRevD.67.064005}{{\em Phys.
  Rev.} {\bfseries D67} (2003) 064005},
  \href{http://arxiv.org/abs/hep-th/0209120}{{\ttfamily arXiv:hep-th/0209120
  [hep-th]}}.

\bibitem{Everett:1957hd}
H.~Everett, ``{Relative state formulation of quantum mechanics},'' {\em Rev.
  Mod. Phys.} {\bfseries 29} (1957) 454.

\bibitem{Schlosshauer:2003zy}
M.~Schlosshauer, ``{Decoherence, the measurement problem, and interpretations
  of quantum mechanics},''
  \href{http://dx.doi.org/10.1103/RevModPhys.76.1267}{{\em Rev. Mod. Phys.}
  {\bfseries 76} (2004) 1267},
  \href{http://arxiv.org/abs/quant-ph/0312059}{{\ttfamily
  arXiv:quant-ph/0312059 [quant-ph]}}.

\bibitem{wallace}
D.~Wallace, {\em The Emergent Multiverse}.
\newblock Oxford University Press, Oxford, 2012.

\bibitem{Elby:1994}
A.~{Elby} and J.~{Bub}, ``{Triorthogonal uniqueness theorem and its relevance
  to the interpretation of quantum mechanics},''
  \href{http://dx.doi.org/10.1103/PhysRevA.49.4213}{{\em Phys. Rev.} {\bfseries
  A49} (1994) 4213}.

\bibitem{Zurek:1981xq}
W.~Zurek, ``{Pointer Basis of Quantum Apparatus: Into What Mixture Does the
  Wave Packet Collapse?},''
  \href{http://dx.doi.org/10.1103/PhysRevD.24.1516}{{\em Phys. Rev.} {\bfseries
  D24} (1981) 1516}.

\bibitem{Zurek:1993ptp}
W.~H. Zurek, ``Preferred states, predictability, classicality and the
  environment-induced decoherence,''
  \href{http://dx.doi.org/10.1143/PTP.89.281}{{\em Prog.~Theor.~Phys.}
  {\bfseries 89} (1993) 281}.

\bibitem{Zurek:1998ji}
W.~H. Zurek, ``{Decoherence, Einselection, and the existential interpretation:
  The Rough guide},'' \href{http://dx.doi.org/10.1098/rsta.1998.0250}{{\em
  Phil. Trans. Roy. Soc. Lond.} {\bfseries A356} (1998) 1793},
  \href{http://arxiv.org/abs/quant-ph/9805065}{{\ttfamily
  arXiv:quant-ph/9805065 [quant-ph]}}.

\bibitem{Zurek:2003rmp}
W.~H. {Zurek}, ``{Decoherence, einselection, and the quantum origins of the
  classical},'' \href{http://dx.doi.org/10.1103/RevModPhys.75.715}{{\em
  Rev.~Mod.~Phys.} {\bfseries 75} (2003) 715},
  \href{http://arxiv.org/abs/arXiv:quant-ph/0105127}{{\ttfamily
  arXiv:quant-ph/0105127}}.

\bibitem{Khlebnikov:2013cmsm}
S.~{Khlebnikov} and M.~{Kruczenski}, ``{Thermalization of isolated quantum
  systems},'' \href{http://arxiv.org/abs/1312.4612}{{\ttfamily arXiv:1312.4612
  [cond-mat.stat-mech]}}.

\bibitem{Page:1983uc}
D.~N. Page and W.~K. Wootters, ``Evolution without evolution: Dynamics
  described by stationary observables,''
  \href{http://dx.doi.org/10.1103/PhysRevD.27.2885}{{\em Phys. Rev.} {\bfseries
  D27} (1983) 2885}.

\bibitem{Page:2005ur}
D.~N. Page, ``{The Lifetime of the universe},'' {\em J. Korean Phys. Soc.}
  {\bfseries 49} (2006) 711,
  \href{http://arxiv.org/abs/hep-th/0510003}{{\ttfamily arXiv:hep-th/0510003
  [hep-th]}}.

\bibitem{Davenport:2010jy}
M.~Davenport and K.~D. Olum, ``{Are there Boltzmann brains in the vacuum?},''
\href{http://arxiv.org/abs/1008.0808}{{\ttfamily arXiv:1008.0808 [hep-th]}}.

\bibitem{Lindblad:1975ef}
G.~Lindblad, ``{On the Generators of Quantum Dynamical Semigroups},''
\href{http://dx.doi.org/10.1007/BF01608499}{{\em Commun. Math. Phys.}
  {\bfseries 48} (1976) 119}.

\bibitem{2008arXiv0809.4403M}
F.~{Marquardt} and A.~{P{\"u}ttmann}, ``{Introduction to dissipation and
  decoherence in quantum systems},''
  \href{http://arxiv.org/abs/0809.4403}{{\ttfamily arXiv:0809.4403
  [quant-ph]}}.

\bibitem{Zurek:1982ii}
W.~Zurek, ``{Environment induced superselection rules},''
\href{http://dx.doi.org/10.1103/PhysRevD.26.1862}{{\em Phys. Rev.} {\bfseries
  D26} (1982) 1862}.

\bibitem{Birrell:1982ix}
N.~D. Birrell and P.~C.~W. Davies, {\em Quantum Fields in Curved Space}.
\newblock Cambridge Monographs on Mathematical Physics. Cambridge University
  Press,
1984.
\newblock

\bibitem{Allen:1985ux}
B.~Allen, ``{Vacuum States in de Sitter Space},''
  \href{http://dx.doi.org/10.1103/PhysRevD.32.3136}{{\em Phys. Rev.} {\bfseries
  D32} (1985) 3136}.

\bibitem{Page:2012fn}
D.~N. Page and X.~Wu, ``{Massless Scalar Field Vacuum in de Sitter
  Spacetime},'' {\em JCAP} {\bfseries 11} (2012) 51,
  \href{http://arxiv.org/abs/1204.4462}{{\ttfamily arXiv:1204.4462 [hep-th]}}.

\bibitem{Candelas:1975du}
P.~Candelas and D.~Raine, ``{General Relativistic Quantum Field Theory-An
  Exactly Soluble Model},''
  \href{http://dx.doi.org/10.1103/PhysRevD.12.965}{{\em Phys. Rev.} {\bfseries
  D12} (1975) 965}.

\bibitem{Geheniau:1968bcs}
J.~G\'eh\'eniau and C.~Schomblond, ``Fonctions de green dans l'univers de de
  sitter,'' {\em Acad. R. Belg. Bull. Cl. Sci.} {\bfseries 54} (1968) 1147.

\bibitem{Schomblond:1976xc}
C.~Schomblond and P.~Spindel, ``{Unicity Conditions of the Scalar Field
  Propagator Delta(1) (x,y) in de Sitter Universe},'' {\em Annales Poincare
  Phys.Theor.} {\bfseries 25} (1976) 67.

\bibitem{Chernikov:1968zm}
N.~Chernikov and E.~Tagirov, ``{Quantum theory of scalar fields in de Sitter
  space-time},'' {\em Annales Poincare Phys.Theor.} {\bfseries A9} (1968) 109.

\bibitem{Tagirov:1972vv}
E.~Tagirov, ``{Consequences of field quantization in de Sitter type
  cosmological models},''
  \href{http://dx.doi.org/10.1016/0003-4916(73)90047-X}{{\em Annals Phys.}
  {\bfseries 76} (1973) 561}.

\bibitem{Mottola:1984ar}
E.~Mottola, ``{Particle Creation in de Sitter Space},''
  \href{http://dx.doi.org/10.1103/PhysRevD.31.754}{{\em Phys. Rev.} {\bfseries
  D31} (1985) 754}.

\bibitem{Bousso:2001mw}
R.~Bousso, A.~Maloney, and A.~Strominger, ``Conformal vacua and entropy in de
  sitter space,'' {\em Phys. Rev.} {\bfseries D65} (2002) 104039.

\bibitem{Anderson:2000wx}
P.~R. Anderson, W.~Eaker, S.~Habib, C.~Molina-Paris, and E.~Mottola,
  ``{Attractor states and infrared scaling in de Sitter space},'' {\em Phys.
  Rev.} {\bfseries D62} (2000) 124019,
  \href{http://arxiv.org/abs/gr-qc/0005102}{{\ttfamily arXiv:gr-qc/0005102
  [gr-qc]}}.

\bibitem{Hollands:2011we}
S.~Hollands, ``{Massless interacting quantum fields in deSitter spacetime},''
  \href{http://dx.doi.org/10.1007/s00023-011-0140-1}{{\em Annales Henri
  Poincare} {\bfseries 13} (2012) 1039},
\href{http://arxiv.org/abs/1105.1996}{{\ttfamily arXiv:1105.1996 [gr-qc]}}.

\bibitem{Garbrecht:2011gu}
B.~Garbrecht and G.~Rigopoulos, ``{Self Regulation of Infrared Correlations for
  Massless Scalar Fields during Inflation},'' {\em Phys. Rev.} {\bfseries D84}
  (2011) 063516, \href{http://arxiv.org/abs/1105.0418}{{\ttfamily
  arXiv:1105.0418 [hep-th]}}.

\bibitem{Garbrecht:2013coa}
B.~Garbrecht, G.~Rigopoulos, and Y.~Zhu, ``{Infrared Correlations in de Sitter
  Space: Field Theoretic vs. Stochastic Approach},'' {\em Phys. Rev.}
  {\bfseries D89} (2014) 063506,
  \href{http://arxiv.org/abs/1310.0367}{{\ttfamily arXiv:1310.0367 [hep-th]}}.

\bibitem{Nomura:2011dt}
Y.~Nomura, ``{Physical Theories, Eternal Inflation, and Quantum Universe},''
  \href{http://dx.doi.org/10.1007/JHEP11(2011)063}{{\em JHEP} {\bfseries 11}
  (2011) 063},
\href{http://arxiv.org/abs/1104.2324}{{\ttfamily arXiv:1104.2324 [hep-th]}}.

\bibitem{Nomura:2011rb}
Y.~Nomura, ``{Quantum Mechanics, Spacetime Locality, and Gravity},''
  \href{http://dx.doi.org/10.1007/s10701-013-9729-1}{{\em Found. Phys.}
  {\bfseries 43} (2013) 978},
\href{http://arxiv.org/abs/1110.4630}{{\ttfamily arXiv:1110.4630 [hep-th]}}.

\bibitem{Bekenstein:1973ur}
J.~D. Bekenstein, ``{Black holes and entropy},''
  \href{http://dx.doi.org/10.1103/PhysRevD.7.2333}{{\em Phys. Rev.} {\bfseries
  D7} (1973) 2333}.

\bibitem{Hawking:1971tu}
S.~Hawking, ``{Gravitational radiation from colliding black holes},''
  \href{http://dx.doi.org/10.1103/PhysRevLett.26.1344}{{\em Phys. Rev. Lett.}
  {\bfseries 26} (1971) 1344}.

\bibitem{Goheer:2002vf}
N.~Goheer, M.~Kleban, and L.~Susskind, ``{The Trouble with de Sitter space},''
  \href{http://dx.doi.org/10.1088/1126-6708/2003/07/056}{{\em JHEP} {\bfseries
  07} (2003) 056},
\href{http://arxiv.org/abs/hep-th/0212209}{{\ttfamily arXiv:hep-th/0212209
  [hep-th]}}.

\bibitem{Banks:2005bm}
T.~Banks, ``{Some thoughts on the quantum theory of stable de Sitter space},''
  \href{http://arxiv.org/abs/hep-th/0503066}{{\ttfamily arXiv:hep-th/0503066
  [hep-th]}}.

\bibitem{Giddings:2007nu}
S.~B. Giddings and D.~Marolf, ``{A Global picture of quantum de Sitter
  space},'' \href{http://dx.doi.org/10.1103/PhysRevD.76.064023}{{\em Phys.
  Rev.} {\bfseries D76} (2007) 064023},
\href{http://arxiv.org/abs/0705.1178}{{\ttfamily arXiv:0705.1178 [hep-th]}}.

\bibitem{Coleman:1977py}
S.~R. Coleman, ``{The Fate of the False Vacuum. 1. Semiclassical Theory},''
  \href{http://dx.doi.org/10.1103/PhysRevD.15.2929,
  10.1103/PhysRevD.16.1248}{{\em Phys. Rev.} {\bfseries D15} (1977) 2929}.

\bibitem{Coleman:1980aw}
S.~R. Coleman and F.~De~Luccia, ``{Gravitational Effects on and of Vacuum
  Decay},'' \href{http://dx.doi.org/10.1103/PhysRevD.21.3305}{{\em Phys.~Rev.}
  {\bfseries D21} (1980) 3305}.

\bibitem{Susskind:2007pv}
L.~Susskind, ``{The Census taker's hat},''
  \href{http://arxiv.org/abs/0710.1129}{{\ttfamily arXiv:0710.1129 [hep-th]}}.

\bibitem{Sekino:2009kv}
Y.~Sekino and L.~Susskind, ``{Census Taking in the Hat: FRW/CFT Duality},''
  \href{http://dx.doi.org/10.1103/PhysRevD.80.083531}{{\em Phys. Rev.}
  {\bfseries D80} (2009) 083531},
\href{http://arxiv.org/abs/0908.3844}{{\ttfamily arXiv:0908.3844 [hep-th]}}.

\bibitem{Bousso:2011up}
R.~Bousso and L.~Susskind, ``{The Multiverse Interpretation of Quantum
  Mechanics},'' \href{http://dx.doi.org/10.1103/PhysRevD.85.045007}{{\em
  Phys.~Rev.} {\bfseries D85} (2012) 045007},
\href{http://arxiv.org/abs/1105.3796}{{\ttfamily arXiv:1105.3796 [hep-th]}}.

\bibitem{Page:2006dt}
D.~N. Page, ``{Is our universe likely to decay within 20 billion years?},''
  \href{http://dx.doi.org/10.1103/PhysRevD.78.063535}{{\em Phys. Rev.}
  {\bfseries D78} (2008) 063535},
  \href{http://arxiv.org/abs/hep-th/0610079}{{\ttfamily arXiv:hep-th/0610079
  [hep-th]}}.

\bibitem{Page:2006hr}
D.~N. Page, ``{Susskind's challenge to the Hartle-Hawking no-boundary proposal
  and possible resolutions},'' {\em JCAP} {\bfseries 1} (2007) 004,
  \href{http://arxiv.org/abs/hep-th/0610199}{{\ttfamily arXiv:hep-th/0610199
  [hep-th]}}.

\bibitem{Page:2006ys}
D.~N. Page, ``{Return of the Boltzmann Brains},''
  \href{http://dx.doi.org/10.1103/PhysRevD.78.063536}{{\em Phys. Rev.}
  {\bfseries D78} (2008) 063536},
  \href{http://arxiv.org/abs/hep-th/0611158}{{\ttfamily arXiv:hep-th/0611158
  [hep-th]}}.

\bibitem{Page:2006nt}
D.~N. Page, ``{Is our universe decaying at an astronomical rate?},''
  \href{http://dx.doi.org/10.1016/j.physletb.2008.08.039}{{\em Phys. Lett.}
  {\bfseries B669} (2008) 197},
  \href{http://arxiv.org/abs/hep-th/0612137}{{\ttfamily arXiv:hep-th/0612137
  [hep-th]}}.

\bibitem{Page:2009mc}
D.~N. Page, ``{Possible Anthropic Support for a Decaying Universe: A Cosmic
  Doomsday Argument},'' \href{http://arxiv.org/abs/0907.4153}{{\ttfamily
  arXiv:0907.4153 [hep-th]}}.

\bibitem{Gott:2008ii}
J.~R. Gott~III, ``{Boltzmann Brains: I'd Rather See Than Be One},''
\href{http://arxiv.org/abs/0802.0233}{{\ttfamily arXiv:0802.0233 [gr-qc]}}.

\bibitem{Aaronson:2013ema}
S.~Aaronson, ``{The Ghost in the Quantum Turing Machine},''
\href{http://arxiv.org/abs/1306.0159}{{\ttfamily arXiv:1306.0159 [quant-ph]}}.

\bibitem{Carroll:2008yd}
S.~M. Carroll, ``{What if Time Really Exists?},''
\href{http://arxiv.org/abs/0811.3772}{{\ttfamily arXiv:0811.3772 [gr-qc]}}.

\bibitem{Boddy:2013qma}
K.~K. Boddy and S.~M. Carroll, ``{Can the Higgs Boson Save Us From the Menace
  of the Boltzmann Brains?},''
\href{http://arxiv.org/abs/1308.4686}{{\ttfamily arXiv:1308.4686 [hep-ph]}}.

\bibitem{Garriga:1997ef}
J.~Garriga and A.~Vilenkin, ``{Recycling universe},''
  \href{http://dx.doi.org/10.1103/PhysRevD.57.2230}{{\em Phys. Rev.} {\bfseries
  D57} (1998) 2230}, \href{http://arxiv.org/abs/astro-ph/9707292}{{\ttfamily
  arXiv:astro-ph/9707292 [astro-ph]}}.

\bibitem{Linde:2006nw}
A.~D. Linde, ``{Sinks in the Landscape, Boltzmann Brains, and the Cosmological
  Constant Problem},''
  \href{http://dx.doi.org/10.1088/1475-7516/2007/01/022}{{\em JCAP} {\bfseries
  1} (2007) 022}, \href{http://arxiv.org/abs/hep-th/0611043}{{\ttfamily
  arXiv:hep-th/0611043 [hep-th]}}.

\bibitem{Polarski:1995jg}
D.~Polarski and A.~A. Starobinsky, ``{Semiclassicality and decoherence of
  cosmological perturbations},''
  \href{http://dx.doi.org/10.1088/0264-9381/13/3/006}{{\em Class. Quant. Grav.}
  {\bfseries 13} (1996) 377},
\href{http://arxiv.org/abs/gr-qc/9504030}{{\ttfamily arXiv:gr-qc/9504030
  [gr-qc]}}.

\bibitem{Lombardo:2005iz}
F.~C. Lombardo and D.~Lopez~Nacir, ``{Decoherence during inflation: The
  Generation of classical inhomogeneities},''
  \href{http://dx.doi.org/10.1103/PhysRevD.72.063506}{{\em Phys. Rev.}
  {\bfseries D72} (2005) 063506},
\href{http://arxiv.org/abs/gr-qc/0506051}{{\ttfamily arXiv:gr-qc/0506051
  [gr-qc]}}.

\bibitem{Martineau:2006ki}
P.~Martineau, ``{On the decoherence of primordial fluctuations during
  inflation},'' \href{http://dx.doi.org/10.1088/0264-9381/24/23/006}{{\em
  Class. Quant. Grav.} {\bfseries 24} (2007) 5817},
\href{http://arxiv.org/abs/astro-ph/0601134}{{\ttfamily arXiv:astro-ph/0601134
  [astro-ph]}}.

\bibitem{Burgess:2006jn}
C.~P. Burgess, R.~Holman, and D.~Hoover, ``{Decoherence of inflationary
  primordial fluctuations},''
  \href{http://dx.doi.org/10.1103/PhysRevD.77.063534}{{\em Phys. Rev.}
  {\bfseries D77} (2008) 063534},
\href{http://arxiv.org/abs/astro-ph/0601646}{{\ttfamily arXiv:astro-ph/0601646
  [astro-ph]}}.

\bibitem{Kiefer:2006je}
C.~Kiefer, I.~Lohmar, D.~Polarski, and A.~A. Starobinsky, ``{Pointer states for
  primordial fluctuations in inflationary cosmology},''
  \href{http://dx.doi.org/10.1088/0264-9381/24/7/002}{{\em Class. Quant. Grav.}
  {\bfseries 24} (2007) 1699},
\href{http://arxiv.org/abs/astro-ph/0610700}{{\ttfamily arXiv:astro-ph/0610700
  [astro-ph]}}.

\bibitem{Prokopec:2006fc}
T.~Prokopec and G.~I. Rigopoulos, ``{Decoherence from Isocurvature
  perturbations in Inflation},'' {\em JCAP} {\bfseries 11} (2007) 029,
  \href{http://arxiv.org/abs/astro-ph/0612067}{{\ttfamily
  arXiv:astro-ph/0612067 [astro-ph]}}.

\bibitem{Creminelli:2008es}
P.~Creminelli, S.~Dubovsky, A.~Nicolis, L.~Senatore, and M.~Zaldarriaga, ``{The
  Phase Transition to Slow-roll Eternal Inflation},''
  \href{http://dx.doi.org/10.1088/1126-6708/2008/09/036}{{\em JHEP} {\bfseries
  09} (2008) 036},
\href{http://arxiv.org/abs/0802.1067}{{\ttfamily arXiv:0802.1067 [hep-th]}}.

\bibitem{Dubovsky:2011uy}
S.~Dubovsky, L.~Senatore, and G.~Villadoro, ``{Universality of the Volume Bound
  in Slow-Roll Eternal Inflation},''
  \href{http://dx.doi.org/10.1007/JHEP05(2012)035}{{\em JHEP} {\bfseries 05}
  (2012) 035},
\href{http://arxiv.org/abs/1111.1725}{{\ttfamily arXiv:1111.1725 [hep-th]}}.

\bibitem{Martinec:2014uva}
E.~J. Martinec and W.~E. Moore, ``{Modeling Quantum Gravity Effects in
  Inflation},''
\href{http://arxiv.org/abs/1401.7681}{{\ttfamily arXiv:1401.7681 [hep-th]}}.

\bibitem{Ade:2014xna}
{\bfseries BICEP2} Collaboration, P.~Ade {\em et~al.}, ``{BICEP2 I: Detection
  Of B-mode Polarization at Degree Angular Scales},''
  \href{http://dx.doi.org/10.1103/PhysRevLett.112.241101}{{\em Phys. Rev.
  Lett.} {\bfseries 112} (2014) 241101},
  \href{http://arxiv.org/abs/1403.3985}{{\ttfamily arXiv:1403.3985
  [astro-ph.CO]}}.

\bibitem{Bousso:1998bn}
R.~Bousso, ``{Proliferation of de Sitter space},''
  \href{http://dx.doi.org/10.1103/PhysRevD.58.083511}{{\em Phys. Rev.}
  {\bfseries D58} (1998) 083511},
\href{http://arxiv.org/abs/hep-th/9805081}{{\ttfamily arXiv:hep-th/9805081
  [hep-th]}}.

\bibitem{Bohm:1951xw}
D.~{Bohm}, ``{A Suggested Interpretation of the Quantum Theory in Terms of
  `Hidden' Variables. I},''
  \href{http://dx.doi.org/10.1103/PhysRev.85.166}{{\em Phys. Rev.} {\bfseries
  85} (1952) 166}.

\bibitem{Bohm:1951xx}
D.~Bohm, ``{A Suggested Interpretation of the Quantum Theory in Terms of
  `Hidden' Variables. II},''
  \href{http://dx.doi.org/10.1103/PhysRev.85.180}{{\em Phys. Rev.} {\bfseries
  85} (1952) 180}.

\bibitem{Durr:2003gu}
D.~D\"urr, S.~Goldstein, R.~Tumulka, and N.~Zangh\`\i, ``Bohmian mechanics and
  quantum field theory,''
  \href{http://dx.doi.org/10.1103/PhysRevLett.93.090402}{{\em Phys. Rev. Lett.}
  {\bfseries 93} (2004) 090402}.

\bibitem{Struyve:2007fs}
W.~Struyve, ``{Pilot-wave theory and quantum fields},''
  \href{http://dx.doi.org/10.1088/0034-4885/73/10/106001}{{\em Rept. Prog.
  Phys.} {\bfseries 73} (2010) 106001},
\href{http://arxiv.org/abs/0707.3685}{{\ttfamily arXiv:0707.3685 [quant-ph]}}.

\bibitem{Goldstein:2015mha}
S.~Goldstein, W.~Struyve, and R.~Tumulka, ``{The Bohmian Approach to the
  Problems of Cosmological Quantum Fluctuations},''
\href{http://arxiv.org/abs/1508.01017}{{\ttfamily arXiv:1508.01017 [gr-qc]}}.

\bibitem{grw85}
G.~Ghirardi, A.~Rimini, and T.~Weber, ``A model for a unified quantum
  description of macroscopic and microscopic systems,'' in {\em Quantum
  Probability and Applications II}, pp.~223--232.
\newblock Springer, 1985.

\bibitem{Ghirardi:1985mt}
G.~C. Ghirardi, A.~Rimini, and T.~Weber, ``Unified dynamics for microscopic and
  macroscopic systems,'' \href{http://dx.doi.org/10.1103/PhysRevD.34.470}{{\em
  Phys. Rev.} {\bfseries D34} (1986) 470}.

\end{thebibliography}\endgroup
\bibliographystyle{utphys}

\end{document}